\documentclass[twocolumn]{aastex63}

\newcommand\mystrut[1][2]{%
    \setlength\mylena{#1\ht\@arstrutbox}%
    \setlength\mylenb{#1\dp\@arstrutbox}%
    \rule[\mylenb]{0pt}{\mylena}}
\usepackage{booktabs}
\usepackage{CJK}
\usepackage{soul}
\usepackage{float}
\makeatletter
\newcommand\notsotiny{\@setfontsize\notsotiny\@vipt\@viipt}
\makeatother

\newcommand{\lowq}{low-$q$ }
\newcommand{\tblspace}{\,\,\,\,\,\,\,\,\,\,\,\,\,\,\,\,\,}

\shorttitle{Near-Sun Asteroids}
\shortauthors{Holt et al.}
\graphicspath{{./}{figures/}}

\begin{document}
\begin{CJK*}{UTF8}{gbsn}

\title{Surface Properties of Near-Sun Asteroids}

\correspondingauthor{Carrie E. Holt}
\email{carrieholt@astro.umd.edu}

\author[0000-0002-4043-6445]{Carrie E. Holt}
\affiliation{Department of Astronomy, University of Maryland, College Park, MD 20742, USA}

\author[0000-0003-2781-6897]{Matthew M. Knight}
\affiliation{Department of Physics, United States Naval Academy, 572C Holloway Rd, Annapolis, MD 21402, USA}
\affiliation{Department of Astronomy, University of Maryland, College Park, MD 20742, USA}

\author[0000-0002-6702-7676]{Michael S.~P. Kelley}
\affiliation{Department of Astronomy, University of Maryland, College Park, MD 20742, USA}

\author[0000-0002-4838-7676]{Quanzhi Ye (叶泉志)}
\affiliation{Department of Astronomy, University of Maryland, College Park, MD 20742, USA}

\author[0000-0001-7225-9271]{Henry H. Hsieh}
\affiliation{Planetary Science Institute, Tucson, AZ 85719, USA} 
\affiliation{Institute of Astronomy and Astrophysics, Academia Sinica, Taipei 10617, Taiwan}

\author[0000-0001-9328-2905]{Colin Snodgrass}
\affiliation{Institute for Astronomy, University of Edinburgh, Royal Observatory, Edinburgh, EH9 3HJ, UK}

\author[0000-0003-0250-9911]{Alan Fitzsimmons}
\affiliation{Astrophysics Research Centre, School of Mathematics and Physics, Queen's University Belfast, Belfast, BT7 1NN, UK}

\author[0000-0002-0054-6850]{Derek C. Richardson}
\affiliation{Department of Astronomy, University of Maryland, College Park, MD 20742, USA}

\author[0000-0002-9413-8785]{Jessica M. Sunshine}
\affiliation{Department of Astronomy, University of Maryland, College Park, MD 20742, USA}

\author[0000-0002-9138-9028]{Nora L. Eisner}
\affiliation{Department of Physics, University of Oxford, Keble Road, Oxford OX1 3RH, UK}

\author[0000-0002-7600-4652]{Annika Gustaffson}
\affiliation{Southwest Research Institute, Boulder, CO 80302, USA}



\begin{abstract}

Near-Earth Asteroids (NEAs) with small perihelion distances reach sub-solar temperatures of $\geq1000$ K. They are hypothesized to undergo ``super-catastrophic" disruption, potentially caused by near-Sun processes such as thermal cracking, spin-up, meteoroid impacts, and subsurface volatile release; all of which are likely to cause surface alteration, which may change the spectral slope of the surface. We attempted to observe 35 of the 53 known near-Sun asteroids with $q\leq0.15$ au from January 2017 to March 2020 to search for trends related to near-Sun processes. We report the optical colors and spectral slopes of 22 objects that we successfully observed and the measured rotation periods for three objects. We find the distribution of colors to be overall bluer than the color distribution of NEAs, though there is large overlap. We attribute large scatter to unknown dynamical histories and compositions for individual objects, as well as competing surface altering processes. We also investigated potential correlations between colors and other properties (e.g., perihelion distance, Tisserand parameter, rotation period), and searched for evidence of activity. Finally, we have compiled all known physical and dynamical properties of these objects, including probabilistic source regions and dwell times with $q\leq0.15$ au.

\end{abstract}


\keywords{Near-Earth objects (1092) --- Solar system astronomy (1529) --- Asteroids (72) --- Solar system astronomy (1529) --- Ground-based astronomy (686) --- Optical astronomy (1776)}


\section{INTRODUCTION}
\label{sec:intro}
\subsection{Background}
\label{sec:background}
Many comets and asteroids spend part of their dynamical lifetimes with small perihelion distances (or ``low-$q$'') as a result of dynamical interactions with Jupiter \citep[e.g.,][]{bailey92b,farinella94,gladman97,bottke02,marchi09,greenstreet12}. However, even while assuming a loss due to collisions with the Sun or planets, or an escape from the inner solar system, \citet{granvik16} reported a deficit of asteroids with small perihelion distances ($q\lesssim0.5$ au), which they attributed to ``super-catastrophic disruption'' on timescales of less than 250 years when an asteroid reaches a perihelion distance less than $q \sim 0.076$ au, with some dependence on asteroid size. Based on the evolution of debris streams, \citet{ye19} suggest that disruption is a more gradual process with timescales of $1-20$ kyr. \citet{granvik16} determined that the disruption of near-Sun asteroids cannot be explained by tidal effects or thermal vaporization. Potential near-Sun processes that might lead to disruption over time (e.g., thermal cracking, spin-up, subsurface volatile release) likely cause surface alteration, which  might be observable from the ground using optical telescopes.

\subsection{Observed Trends}

Low-$q$ comets 322P/SOHO 1, 96P/Machholz 1 and 323P/SOHO have been observed to have atypical blue colors \citep{knight16,eisner19,hui22}. 322P/SOHO 1 has a cometary orbit and shows evidence of being active at perihelion \citep{lamy13}, but it has an asteroid-like albedo, density, and fast rotation. 96P/Machholz 1 had the smallest perihelion distance of any unambiguously active short-period comet until recent observations of 323P/SOHO, which was observed post-perihelion with a long-narrow tail, likely caused by immense thermal stress or rotational instability; 323P has the shortest rotational period of known comets in our Solar System at 0.522 hrs. 323P/SOHO 1 was recently observed to have bluer colors than most Solar System small bodies, which changed unprecedentedly over a few weeks to a color unlike any other small bodies in our Solar System \citep{hui22}. Further characterization of objects with small perihelion distances is necessary to assess whether these comets are unique or if their bluer colors are typical of objects that closely approach the Sun. Very few periodic comets are observed from the ground with a perihelion distance less than 0.15 au, so additional observations must focus on asteroids.

Spectral trends with perihelion distance have been observed for S- and Q-type near-Earth asteroids (NEAs) with $q \gtrsim 0.2$ au. \citet{marchi06b} observed that S- and Q-types become spectrally bluer with decreasing perihelion distance using data from \citet{binzel04} and \citet{lazzarin04, lazzarin05}, which they attribute to resurfacing caused by close encounters with planets. Because of their closely related spectra, only different in features impacted by space weathering (spectral slope and 1-$\mu$m band depth), Q-types are best explained as resurfaced S-types that have not yet had enough time for space weathering to alter their surface \citep[e.g.,][]{chapman04,brunetto15}. In agreement with \citet{marchi06b}, \citet{demeo14} and \citet{devogele19} found a higher percentage of Q-type asteroids (which are spectrally bluer than S-types) with smaller perihelion distances.  \citet{graves19} argue that resurfacing due to thermal degradation explains the trend better than planetary encounters after modeling NEA orbits and tracking spectral slope with planetary encounters or physical distance from the Sun for each object. More observations of near-Sun asteroids are needed to determine if this trend holds at smaller perihelion distances.

\newpage

\subsection{Overview of Known Properties}
\label{sec:previouswork}

This work presents an observational study of known near-Sun asteroids and searches for any common properties that might be related to near-Sun processes experienced by these objects. We focus our efforts on objects with perihelion distance $q \leq 0.15$ au. We chose this limit because 0.15 au is the outer limit of the Solar and Heliospheric Observatory (SOHO)'s outermost coronagraph (C3)'s field-of-view, meaning we can potentially view activity or disruption of these objects using SOHO. There are 53 known asteroids that reach perihelion at $q \leq 0.15$ au (sub-solar temperatures $\gtrsim1000$ K) as of November 1, 2021, including the ones with poorly constrained orbits. All of the objects reach perihelion within the SOHO field of view each orbit, yet none have been observed to date, meaning that mass loss near perihelion has been very low over the last quarter century (to be explored further in Section~\ref{sec:activity}).

\citet{campins09} and \citet{jewitt13b} investigated a similar set of asteroids, but their perihelion distance limits were significantly larger ($q\leq 0.35$ au and $\leq0.25$ au, respectively). Only seven of their objects are in our sample. Most \lowq asteroids are small and therefore faint and difficult to study. In-depth studies have not been possible for \lowq objects other than 3200 Phaethon. While there have been a few serendipitous observations as part of general NEO studies, the available data are still limited, as detailed below and summarized in Table~\ref{tab:published}.

\begin{deluxetable*}{lccccccccc}
\tablenum{1}
\tablecaption{Published Measurements of Near-Sun Asteroids*\label{tab:published}}
\tablecolumns{10}
\tablewidth{0pt} 
\tabletypesize{\scriptsize}
\setlength{\tabcolsep}{0.1in}
\renewcommand{\arraystretch}{1.3}
\tablehead{   
  \colhead{Object} & 
  \colhead{$B - V$}&
  \colhead{$V - R$}&
  \colhead{$B - R$}&
  \colhead{Ref.}&
  \colhead{Rotation Rate [hrs]}& 
  \colhead{Ref.}& 
  \colhead{Albedo}&
  \colhead{Size [km]}&
  \colhead{Ref.}
}
\startdata
1995 CR&-&-&-&-&$2.66\pm0.04$&1&$0.167\pm_{0.079}^{0.101}$&$0.129\pm_{0.024}^{0.045}$&2\\
2000 BD19&-&-&-&-&$10.570\pm0.005$&3&$0.247\pm0.046$&$0.97\pm0.04$&4\\
&&&&&&&$0.123\pm_{0.066}^{0.082}$&$1.149\pm_{0.241}^{0.513}$&5\\
2000 LK&-&-&-&-&-&-&$0.137\pm_{0.074}^{0.092}$&$0.610\pm_{0.132}^{0.278}$&5\\
2002 AJ129&$0.687\pm0.050$&$0.405\pm0.027$&$1.092\pm0.057$&6&$3.9222\pm0.0008$&6&$0.226\pm_{0.116}^{0.141}$&$0.438\pm_{0.091}^{0.181}$&5\\
&-&-&$1.23\pm0.10$&7&-&-&-&-&-\\
2002 PD43&$0.66\pm0.05$&$0.42\pm0.05$&$ 1.08\pm0.03$&7&-&-&-&-&-\\
2004 UL&$0.82\pm0.10$&$0.54\pm0.10$&$1.37\pm0.10$&7&$38\pm2$&8&$0.604\pm_{0.264}^{0.258}$&$0.268\pm_{0.042}^{0.083}$&5\\
2004 XY60&-&-&-&-&-&-&$0.217\pm_{0.110}^{0.137}$&$0.389\pm_{0.080}^{0.161}$&5\\
2006 HY51&-&-&-&-&$3.350\pm0.008$&9&$0.157\pm0.071$&$1.218\pm0.23$&4\\
2006 TC&$0.60\pm0.08$&$0.33\pm0.03$&$0.93\pm0.08$&7&-&-&-&-&-\\
2007 EP88&-&-&-&-&-&-&$0.174\pm0.038$&$0.636\pm0.04$&4\\
&&&&&&&$0.531\pm_{0.230}^{0.236}$&$0.312\pm_{0.051}^{0.098}$&5\\
2008 HE&-&-&-&-&-&-&$0.120\pm_{0.066}^{0.091}$&$0.779\pm_{0.184}^{0.375}$&10\\
2008 MG1&-&-&-&-&-&-&$0.431\pm_{0.223}^{0.252}$&$0.194\pm_{0.038}^{0.084}$&10\\
2008 XM&-&-&-&-&-&-&$0.128\pm0.032$&$0.367\pm0.01$&4\\
2010 JG87&-&-&-&-&-&-&$0.202\pm0.040$&$0.408\pm0.02$&4\\
2011 KE&-&-&-&-&$8.0\pm0.1$&11&-&-&-\\
2011 XA3&-&$0.473\pm 0.051$**&-&12&$0.730\pm0.007$&12&$0.347\pm_{0.163}^{0.196}$&$0.163\pm_{0.029}^{0.056}$&10\\
2017 AF5&-&-&-&-&$49.68\pm0.06$&13&-&-&-\\
3200 Phaethon&$0.67\pm0.02$&$0.32\pm0.02$&$0.99\pm0.02$&7&$3.6039\pm0.0002$&14&$0.16\pm0.02$&$4.6\pm0.3$&15\\
\enddata
\tablerefs{(1) \citet{mpb2014july}
; (2) \citet{trilling16_survey} (3) \citet{mpb2015july}; (4) \citet{mainzer11,mainzer16}; (5) \citet{trilling10}; (6) \citet{devyatkin22}; (7) \citet{jewitt13}; (8) \citet{mpb2015april}; (9) This work; (10) \citet{trilling16_legacy}; (11) Skiff, B.A. (2011)\textsuperscript{a}(12) \citet{urakawa14}; (13) \citet{mpb2017july}; (14) \citet{mpb2015april};(15) \citet{masiero19}\\
\text{*} All uncertainty values are presumed 1-$\sigma$ errors\\
\text{**} Converted from $g'-r'$ according to \citet{jordi06}\\
\text{\textsuperscript{a}} Posting on CALL web site. http://www.minorplanet.info/call.html}
\end{deluxetable*}

\subsubsection{3200 Phaethon}
Phaethon is the most studied \lowq asteroid because it is the largest among the sample (diameter $\sim$5 km) and has a short orbital period (1.43 yr), allowing for observations nearly every apparition, including a historically close approach to Earth in 2017. The parent body of the Geminid meteor shower \citep{whipple83}, Phaethon is the only known unambiguously active \lowq asteroid \citep{jewitt10,jewitt13,li13b,hui17} and the only named asteroid that approaches within 0.15 au of the Sun. Phaethon is a B-type asteroid \citep{binzel01} with an albedo of $\sim$0.16 \citep{masiero19} and a rotational period of $\sim$3.6 hours \citep{mpb2015april}. Spectroscopy and colors have been measured extensively (most recently by \citealt{lin20}). The Japanese space agency JAXA's forthcoming DESTINY\textsuperscript{+} mission is expected to fly by Phaethon in 2024 \citep{ozaki22}.

\subsubsection{Albedos}
Near-Earth Object Wide-Field Infrared Survey Explorer (NEOWISE), an infrared characterization survey, has measured diameters and determined albedos for 2000 BD19, 2006 HY51, 2007 EP88, 2008 XM, and 2010 JG87 \citep{mainzer11, mainzer16}. Albedos were derived by combining infrared measurements with previously reported or follow-up optical magnitudes using the Near-Earth Asteroid Thermal Model \citep[NEATM;][]{harris98}. Additional albedo measurements were made using the Spitzer Space Telescope via the ExploreNEOs \citep{trilling10} and NEOSurvey \citep{trilling16_survey} infrared characterization surveys. We include both NEOWISE and Spitzer albedos in Table \ref{tab:published} for completeness, but we caution the use of Spitzer measurements with reported albedos that are larger than the assumed NEO upper limit of $\sim 0.5$ \citep{gustafsson19}.

\subsubsection{Rotation Periods}
Rotation rates have been measured for eight \lowq asteroids, including Phaethon. All objects except 2011 XA3 have rotation rates ranging from $\sim$2.5 hours \citep[1995 CR;][]{mpb2014july} to $\sim$2 days \citep[2017 AF5;][]{mpb2017july}. 2011 XA3 is the exception with a fast rotation rate of $\sim$45 minutes \citep{urakawa14}. We discuss this object further in Section \ref{sec:discussion_rot}. 

\subsubsection{Spectral Properties}
Spectral observations are limited for the \lowq population. Three objects in addition to Phaethon have measured IR spectra and classification: 137924 (2000 BD19) is V type, 394130 (2006 HY51) is R-type, and 465402 (2008 HW1) is S-type \citep{thomas14, binzel19}. The variety of taxonomies is interesting to note considering the majority of NEAs are S-type,  while V- and R- types only make up $\sim5\%$ of the NEA population \citep{binzel19}.

\begin{deluxetable*}{rlcclcrrccccccccccrc}
\tablenum{2}
\tablecaption{Near-Sun Asteroid Properties\textsuperscript{a}\label{tab:orbits}}
\tablecolumns{20}
\tablewidth{0pt} 
\tabletypesize{\notsotiny}
\setlength{\tabcolsep}{0.05in}
\tablehead{   
  \\&&&\multicolumn{4}{c}{Orbital Elements\textsuperscript{b}}&&&&&
  \multicolumn{7}{c}{Source Region Probabilities\textsuperscript{c}}\\
  \cmidrule(r){4-7}\cmidrule(l){12-18}
  \multicolumn{2}{c}{Object} & 
  \colhead{$H_V$\textsuperscript{d}}&
  \colhead{$q$}&
  \colhead{$a$}&
  \colhead{$e$}&
  \colhead{$i$}&
  \colhead{$T_J$}&
  \colhead{U\textsuperscript{e}}&
  \colhead{$T_{BB}(q)$\textsuperscript{f}}&
  \colhead{$T_{SS}(q)$\textsuperscript{g}}&
  \colhead{$\nu_6$}&
  \colhead{5:2}&
  \colhead{2:1}&
  \colhead{Hun}&
  \colhead{3:1}&
  \colhead{Pho}&
  \colhead{JFC}&
  \colhead{$T_{q*}$\textsuperscript{h}}&
  \colhead{Status\textsuperscript{i}}
}
\startdata
&(2005 HC4)&20.7&0.070&1.824&0.961&8.4&3.17&9&1048&1482&\textbf{0.76}&0.04&0.00&0.01&0.18&0.00&0.00&-\textsuperscript{j}&A\\
&(2020 BU13)&21.2&0.073&2.471&0.970&9.2&2.44&6&1028&1454&0.18&0.11&0.00&0.00&\textbf{0.69}&0.00&0.01&-&X\\
&(2017 TC1)&20.9&0.076&2.491&0.970&9.3&2.42&8&1010&1428&0.19&0.13&0.00&0.00&\textbf{0.66}&0.00&0.01&-&O\textsuperscript{k}\\
&(2008 FF5)&23.1&0.080&2.272&0.965&2.6&2.64&9&986&1395&\textbf{0.62}&0.01&0.00&0.02&0.34&0.00&0.02&-&X\\
&(2017 MM7)&21.1&0.080&2.064&0.961&23.2&2.84&7&983&1391&\textbf{0.63}&0.06&0.00&0.02&0.27&0.02&0.00&-&A\\
&(2015 EV)&22.5&0.081&2.036&0.960&11.4&2.90&9&980&1386&\textbf{0.70}&0.02&0.00&0.02&0.26&0.00&0.00&-&X\\
394130&(2006 HY51)&17.1&0.081&2.588&0.969&33.6&2.30&0&975&1379&0.08&0.26&0.01&0.01&\textbf{0.55}&0.06&0.03&6.1&O\\
&(2021 AF3)&22.9&0.086&1.301&0.934&7.2&4.35&9&946&1338&\textbf{0.89}&0.01&0.00&0.05&0.05&0.00&0.00&-&X\\
&(2016 GU2)&24.0&0.087&2.052&0.957&10.3&2.89&8&941&1331&\textbf{0.68}&0.00&0.00&0.04&0.28&0.00&0.00&-&A\\
&(2019 AM13)&22.0&0.091&1.296&0.930&16.7&4.37&2&922&1304&\textbf{0.78}&0.02&0.00&0.06&0.14&0.00&0.00&-&O\\
&(2019 JZ6)&21.0&0.091&2.461&0.963&24.1&2.45&9&921&1303&0.12&0.15&0.01&0.00&\textbf{0.70}&0.01&0.01&-&X\\
137924&(2000 BD19)&17.4&0.092&0.876&0.895&25.7&6.27&0&917&1297&\textbf{0.52}&0.00&0.00&0.33&0.13&0.01&0.00&7351.8&O\\
374158&(2004 UL)&18.7&0.093&1.266&0.927&23.8&4.45&0&913&1291&\textbf{0.80}&0.03&0.00&0.06&0.09&0.02&0.00&157.0&O\\
394392&(2007 EP88)&18.5&0.096&0.837&0.886&20.7&6.56&1&900&1273&\textbf{0.74}&0.00&0.00&0.14&0.11&0.01&0.00&13687.6&O\\
&(2011 KE)&19.8&0.100&2.207&0.955&5.9&2.74&1&879&1243&\textbf{0.75}&0.05&0.00&0.01&0.19&0.00&0.00&4.0&O\\
465402&(2008 HW1)&17.4&0.103&2.587&0.960&10.5&2.40&0&865&1224&0.14&0.24&0.01&0.00&\textbf{0.55}&0.02&0.04&6.7&O\\
&(2015 HG)&21.0&0.105&2.102&0.950&17.7&2.85&9&859&1215&\textbf{0.62}&0.05&0.00&0.01&0.30&0.02&0.00&4.9&A\\
&(2012 US68)&18.3&0.106&2.503&0.958&25.8&2.44&2&856&1210&0.11&0.29&0.03&0.00&\textbf{0.51}&0.04&0.01&3.7&A\textsuperscript{l}\\
&(2011 XA3)&20.4&0.109&1.467&0.926&28.0&3.90&0&844&1193&\textbf{0.76}&0.03&0.00&0.05&0.15&0.02&0.00&1319.4&X\\
399457&(2002 PD43)&19.1&0.109&2.507&0.956&26.0&2.44&0&841&1189&0.11&0.28&0.03&0.00&\textbf{0.53}&0.03&0.01&2.9&O\\
&(2018 GG5)&19.8&0.110&1.986&0.945&16.8&3.01&7&840&1187&\textbf{0.71}&0.03&0.00&0.01&0.23&0.02&0.00&36.4&O\\
386454&(2008 XM)&20.0&0.111&1.222&0.909&5.4&4.66&0&834&1180&\textbf{0.84}&0.03&0.00&0.02&0.11&0.00&0.00&96.9&O\\
431760&(2008 HE)&18.1&0.112&2.262&0.950&9.8&2.70&0&831&1175&\textbf{0.68}&0.08&0.00&0.01&0.21&0.00&0.01&8.4&O\\
&(2020 DD)&23.5&0.116&2.483&0.953&2.3&2.51&8&818&1156&0.22&0.01&0.00&0.01&\textbf{0.68}&0.00&0.08&-&X\\
276033&(2002 AJ129)&18.7&0.116&1.371&0.915&15.5&4.20&0&815&1153&\textbf{0.84}&0.03&0.01&0.05&0.06&0.01&0.00&88.3&O\\
&(2020 GB2)&21.1&0.117&2.338&0.950&15.2&2.63&7&814&1152&0.42&0.09&0.00&0.01&\textbf{0.48}&0.00&0.00&1.1&X\\
&(2019 VE3)&23.3&0.117&1.174&0.901&2.5&4.85&9&814&1151&\textbf{0.70}&0.00&0.00&0.04&0.26&0.00&0.00&-&X\\
425755&(2011 CP4)&21.2&0.118&0.912&0.870&9.5&6.11&0&809&1144&\textbf{0.95}&0.00&0.00&0.02&0.02&0.01&0.00&2111.9&O\\
&(1995 CR)&21.8&0.119&0.907&0.868&4.1&6.15&0&805&1139&\textbf{0.98}&0.00&0.00&0.02&0.00&0.00&0.00&190.2&O\\
&(2000 LK)&18.3&0.121&2.184&0.945&16.6&2.79&0&801&1132&\textbf{0.66}&0.07&0.00&0.02&0.21&0.04&0.00&19.9&O\\
&(2007 GT3)&19.7&0.122&2.006&0.939&25.6&2.98&2&796&1126&\textbf{0.66}&0.09&0.00&0.02&0.19&0.04&0.00&5.8&A\textsuperscript{m}\\
&(2020 HY2)&24.9&0.124&2.314&0.946&11.4&2.67&9&788&1115&0.43&0.00&0.00&0.03&\textbf{0.50}&0.00&0.05&-&X\\
&(2017 AF5)&17.8&0.124&2.480&0.950&20.9&2.50&0&788&1115&0.13&0.27&0.02&0.00&\textbf{0.53}&0.04&0.01&4.5&O\\
&(2004 QX2)&21.7&0.125&1.286&0.903&19.1&4.45&8&786&1112&\textbf{0.82}&0.03&0.00&0.05&0.09&0.00&0.00&16.9&X\\
&(2019 YV2)&21.9&0.126&1.227&0.897&6.5&4.67&9&783&1107&\textbf{0.76}&0.02&0.00&0.03&0.19&0.00&0.00&11.0&X\\
&(2020 TS2)&18.9&0.126&2.509&0.950&20.1&2.48&8&783&1107&0.12&0.28&0.03&0.00&\textbf{0.49}&0.03&0.05&2.9&X\\
&(2021 LM1)&20.4&0.127&2.312&0.945&29.5&2.63&6&780&1103&0.27&0.13&0.00&0.01&\textbf{0.56}&0.03&0.00&2.4&X\\
&(2011 BT59)&20.9&0.130&2.471&0.947&3.6&2.55&9&771&1091&0.33&0.07&0.00&0.00&\textbf{0.54}&0.00&0.04&1.5&A\\
289227&(2004 XY60)&18.9&0.130&0.640&0.797&23.8&8.52&1&771&1090&\textbf{0.88}&0.00&0.00&0.03&0.06&0.03&0.00&622.2&X\\
&(2015 KO120)&22.0&0.130&1.779&0.927&2.1&3.37&9&770&1089&\textbf{0.77}&0.00&0.00&0.02&0.21&0.00&0.00&5.0&A\\
&(2007 PR10)&20.9&0.132&1.232&0.893&20.9&4.63&1&765&1082&\textbf{0.84}&0.04&0.00&0.04&0.07&0.01&0.00&45.2&A\textsuperscript{l}\\
&(2021 PH27)&17.7&0.133&0.462&0.712&31.9&11.62&3&762&1078&\textbf{0.76}&0.00&0.00&0.04&0.19&0.01&0.00&2236.6&X\\
504181&(2006 TC)&18.7&0.136&1.538&0.912&19.6&3.80&1&755&1068&\textbf{0.77}&0.00&0.00&0.06&0.14&0.03&0.00&50.6&O\\
&(2019 UJ12)&22.4&0.137&2.423&0.943&27.5&2.55&8&751&1063&0.08&0.06&0.00&0.00&\textbf{0.85}&0.00&0.01&0.2&O\\
&(2013 JA36)&21.0&0.138&2.665&0.948&42.5&2.29&8&750&1061&0.04&0.36&0.05&0.00&\textbf{0.50}&0.02&0.02&1.7&A\\
&(2008 MG1)&19.9&0.139&0.783&0.823&5.7&7.08&1&746&1056&\textbf{0.93}&0.00&0.00&0.06&0.00&0.01&0.00&1685.4&O\\
&(2013 HK11)&20.7&0.139&2.182&0.936&17.7&2.82&9&745&1053&\textbf{0.64}&0.05&0.00&0.01&0.28&0.02&0.00&6.0&A\\
&(2017 SK10)&21.4&0.140&2.051&0.932&24.5&2.95&9&743&1051&\textbf{0.60}&0.06&0.00&0.02&0.30&0.02&0.00&1.2&X\\
3200 Phaethon&(1983 TB)&14.3&0.140&1.271&0.890&22.3&4.51&0&743&1051&\textbf{0.67}&0.06&0.00&0.18&0.06&0.03&0.00&1106.0&O\\
&(2020 VL4)&18.5&0.142&2.125&0.933&55.8&2.71&8&737&1043&0.18&0.08&0.20&0.03&\textbf{0.38}&0.14&0.00&363.6&X\\
&(2013 YC)&21.4&0.142&2.495&0.943&2.8&2.55&2&737&1043&0.30&0.06&0.00&0.00&\textbf{0.59}&0.00&0.04&1.0&O\\
&(2020 HE)&23.6&0.146&2.515&0.942&20.5&2.51&9&727&1028&0.06&0.01&0.00&0.00&\textbf{0.84}&0.00&0.08&-&X\\
&(2010 JG87)&19.2&0.148&2.768&0.947&16.8&2.33&0&723&1023&0.01&\textbf{0.56}&0.08&0.00&0.17&0.01&0.17&2.1&O\\
\enddata
\tablenotetext{\text{a}} {As of November 1, 2021}
\vspace*{-3mm}
\tablenotetext{\text{b}} {According to JPL Horizons}
\vspace*{-3mm}
\tablenotetext{\text{c}} {Probability that the object escaped from each of the six source regions, totaling one, using the model detailed in \citet{granvik18a}. The source regions consist of the $\nu_6$ inner main-belt region, Jupiter resonance complexes: 3:1, 5:2 and 2:1, the Hungarias, the Phocaeas, and the Jupiter-family comets.}
\vspace*{-3mm}
\tablenotetext{\text{d}} {Absolute $V$ magnitude as reported by JPL Horizons}
\vspace*{-3mm}
\tablenotetext{\text{e}} {MPC orbit uncertainty estimate. The range is 0-9, with 0 being good and 9 being highly uncertain}
\vspace*{-3mm}
\tablenotetext{\text{f}} {Equilibrium isothermal, spherical blackbody temperature at the perihelion distance, K}
\vspace*{-3mm}
\tablenotetext{\text{g}} {Equilibrium sub-solar blackbody temperature at the perihelion distance, K}
\vspace*{-3mm}
\tablenotetext{\text{h}} {Dwell time with $q\leq0.15$ au in kyrs according to \citet{toliou21}}
\vspace*{-3mm}
\tablenotetext{\text{i}} {O: Observed; A: Attempted but not observed; X: Not attempted}
\vspace*{-3mm}
\tablenotetext{\text{j}} {Some cells result in a $T_{q*}=0$ in the \citet{toliou21} model after renormalizing each cell of the model according to the relative fraction of NEOs from each ER, as explained in \citet{toliou21} Sec. 2.4. We use `-' in those instances.}
\vspace*{-3mm}
\tablenotetext{\text{k}} {2017 TC1 was detected, but was observed in the Milky Way. Colors could not be determined because too few images were obtained for effective DIA (see text).}
\vspace*{-3mm}
\tablenotetext{\text{l}} {2007 PR10 and 2012 US68 were attempted under poor weather conditions}
\vspace*{-3mm}
\tablenotetext{\text{m}} {2007 GT3 had an orbit code of 8 at the time of observations, and was outside the telescope's field-of-view.}

\end{deluxetable*}

\citet{campins09} measured the 7-14 $\mu$m thermal emission spectra of 19 asteroids with $q\leq0.35$ au, including two objects with $q\leq0.15$ au: 2000 BD19 and 2004 XY60. They fit the spectra with thermal continuum models to derive effective diameter, geometric albedo, and beaming parameter of these objects and suggest that thermal behavior is different for near-Sun asteroids compared with other NEAs (stronger thermal emission ``beaming'' at low phase angles). \citet{jewitt13b} measured the colors of nine asteroids with $q\leq0.25$ au to search for evidence of thermal modification. Five of the objects have $q\leq0.15$ au and are therefore a part of our sample. They found that near-Sun objects have a wide range of colors, but there was no statistical distinction between such objects and other near-Earth objects. Both \citet{campins09} and \citet{jewitt13b} consider objects beyond our perihelion cutoff of 0.15 au and suggest further characterization is needed.

\subsection{This Work}
Near-Sun asteroids have not been thoroughly examined as a population. Color observations can provide us with a better understanding of the processes occurring and the effects they have on near-Sun asteroids. Broadband optical colors (e.g., $g'- r'$, $r'-i'$) can be obtained quickly and for fainter objects, which makes such observations optimal for a population study.  Over more than three years of observations from January 2017 to March 2020, we attempted to observe as many near-Sun asteroids as possible. Of the 53 known asteroids with $q\leq0.15$ au (summary of orbital elements in Table \ref{tab:orbits}), we attempted to observe 35, and successfully observed 22. Nine of the objects we attempted to observe were predicted to be within our fields of view and above our detection limit but were not recovered, most likely due to the very large uncertainty in their orbits. We also observed near-Earth asteroid 2004 LG, which previously spent 2500 years with a perihelion distance less than 0.076 au, the disruption limit of \citet{granvik16}, experiencing extreme temperatures of $\sim2500$ K at the surface \citep{vokrouhlicky12b, wiegert20}.

\section{OBSERVATIONS AND DATA ANALYSIS}
\label{sec:observations}

\subsection{Data Collection}
\label{sec:sample}
 
Data were collected primarily using Lowell Observatory's 4.3-m Lowell Discovery Telescope (LDT; formerly known as the Discovery Channel Telescope, DCT) and the 4.1-m Southern Astrophysical Research (SOAR) telescope, supplemented by data from the Isaac Newton Telescope (INT) and Lowell Observatory's 42-in and 31-in telescopes. Broadband SDSS {\it g$'$, r$'$, i$'$, z$'$} filters were used at all telescopes except for Lowell Observatory's 31-inch and 42-inch, where Johnson-Cousins {\it B,V,R,I} filters were used. A summary of the instruments used can be found in Table \ref{tab:instruments}. All images were acquired at the asteroid's ephemeris rate, which frequently resulted in trailed stars. Whenever possible, we kept the star trails to less than two times the seeing in order to accurately register the image (see Sec~\ref{sec:analysis}), but for some objects long trails were unavoidable. As will be discussed later, this occasionally hampered absolute calibrations. Exposure times varied with the asteroid's brightness and telescope size, but were typically 180--300 seconds. The number of images acquired for each target varied with time available, though at least two cycles of filters were achieved for all observations, ordered in a way so that the mid-time of the observations for each filter is approximately the same to mitigate rotational variability. 

\begin{deluxetable*}{llccccl}
\tablenum{3}
\tablecaption{Instruments\label{tab:instruments}}
\tablecolumns{7}
\tablewidth{0pt} 
\tabletypesize{\scriptsize}
\setlength{\tabcolsep}{0.1in}
\renewcommand{\arraystretch}{1.3}
\tablehead{   
  \colhead{Telescope} & 
  \colhead{Instrument}&
  \colhead{Aper Size}&
  \colhead{FOV}&
  \colhead{Pix Sc\textsuperscript{a}}&
  \colhead{Binning}&
  \colhead{Filters}
}
\startdata
Lowell Discovery Telescope (LDT)&Large Monolithic Imager (LMI)&4.3m&$12.3'\times12.3'$&$0.36''$&$3\times3$&$g',r',i',z'$\\
Southern Astrophysical Research (SOAR)&Goodman Spectrograph&4.1m&$7.2'$ diameter&$0.30''$&$2\times2$&$g',r',i',z'$\\
Southern Astrophysical Research (SOAR)&SOAR Optical Imager (SOI)&4.1m&$5.26'\times5.26'$&$0.077''$&$2\times2$&$g',r',i',z'$\\
Isaac Newton Telescope (INT)&Wide Field Campera (WFC)&2.5m&$11.5'\times23.0'$&$0.33''$&None&$g',r',i',z'$\\
Lowell Observatory 31-inch &NASAcam&0.8m&$15.7'\times15.7'$&$0.46''$&None&$B,V,R,I$\\
Lowell Observatory 42-inch&NASA42 Camera&1.1m&$25.3'\times25.3'$&$1.48''$&$3\times3$&$B,V,R,I$\\
\enddata
\tablenotetext{\text{a}}{Effective pixel scale after binning}
\end{deluxetable*}

\subsection{Instruments}
\label{sec:instruments}
Northern hemisphere targets were primarily observed with Lowell Observatory's 4.3-m LDT located near Flagstaff, AZ. All observations were made using the Large Monolithic Imager \citep[LMI;][]{massey13}. LMI has a field-of-view (FOV) of 12\farcm3 $\times$ 12\farcm3 and a pixel scale of 0\farcs36 after an on-chip 3 $\times$ 3 binning. SDSS broadband $g'$, $r'$, $i'$, and $z'$ filters were used.

Observations of southern hemisphere targets were made using the 4.1-m SOAR Telescope on Cerro Pach\'{o}n in Chile. The Goodman Spectrograph Red Camera \citep{clemens04} and the SOAR Optical Imager (SOI) were used. The Goodman Spectrograph camera has a circular FOV of 7\farcm2 diameter. The observations were done in 2 $\times$ 2 binning mode with an effective pixel scale of 0\farcs30. SOI uses two adjacent CCD chips read out through two amplifiers per chip that cover a 5\farcm26 $\times$ 5\farcm26 FOV and have a scale of 0\farcs077/pixel after 2 $\times$ 2 binning. Broadband $g'$, $r'$, $i'$, and $z'$ SDSS filters were used on all images.  

We conducted an observing run in January 2017 using the 2.5-m INT at the Rogue de Los Muchachos Observatory in La Palma, Spain. Observations were made using the Wide Field Camera (WFC), which has an effective FOV of 11\farcm5 $\times$ 23\farcm0 and a pixel scale of 0\farcs33. All observations used broadband $g'$, $r'$, $i'$, and $z'$ SDSS filters.

Supplemental observations were made using Lowell Observatory's Hall 31-inch (0.8 m) and 42-inch (1.1 m) telescopes. The 31-inch telescope has a square field of view of 15\farcm7 on a side and an unbinned pixel scale of 0\farcs46. The 42-inch telescope has a square field of view of 25\farcm3 on a side and a 3 $\times$ 3 binned pixel scale of 0\farcs98. Observations using the 31-inch were robotically acquired. All other telescopes used classic observing mode. Johnson-Cousins $B$, $V$, $R$, and $I$ filters were used.

\subsection{Data Reduction}
\label{sec:dataredux}
Each observation followed the same reduction routine. A master bias frame for each night was created by averaging 10--20 individual bias frames. The master bias frame was then subtracted from each frame. At least five sky or dome flats taken during the same observing period were normalized and median combined for each filter used. Images were flat-field corrected by dividing each frame by the median-combined flat-field.

\subsection{Data Analysis}
\label{sec:analysis}
\subsubsection{Absolute Calibration}
Reduced images were registered using {\tt PHOTOMETRYPIPELINE} \citep{mommert17},  which utilizes {\tt SCAMP} \citep{scamp} to match the source catalog of an image created by Source Extractor \citep{sextractor} with astrometric catalogs, such as GAIA. After registration, using {\tt PHOTOMETRYPIPELINE}, photometry was completed on the image sources followed by the derivation of the photometric image zeropoints with the Pan-STARRS DR1 catalog \citep[PS1;][]{magnier13} using field stars. The aperture size was derived using a curve-of-growth analysis, where the optimum aperture radius is the smallest aperture radius where the target and the background fractional fluxes exceed 70\% and the difference between the target and background curves is less than 5\% (reducing systematic offsets in the flux measurements).

We took our calibration one step further by adding a color correction applied image by image using components of {\tt calviacat} \citep{kelley19}. The difference between instrumental magnitudes and absolute magnitudes of stars is positively correlated with star color. Using the same in-field stars from the PS1 catalog, we determined the color-correction coefficient. We used the previously determined color to calculate the new absolute magnitude. We repeated this process iteratively until the color converged, usually about three times. 

The zeropoints were then converted to either the SDSS photometry system or \textit{BVRI} system depending on the filters used \citep{tonry12}. A weighted average of all object magnitudes of the same filter was measured before calculating the difference to determine the object colors. Colors determined in \textit{BVRI} filters were converted to SDSS colors according to \citet{jordi06}.

As previously discussed, observations were not always acquired during ideal epochs. Therefore, two of the observations (2008 XM and 2018 GG5) include large star streaks ($\sim15''$) as the telescope tracked the asteroid during long exposures required to achieve necessary signal-to-noise. For those observations, we derived the zeropoint from field stars chosen manually, but still using the PS1 catalog.

Images acquired using the 31-inch telescope did not contain a sufficient amount of stars for in-field calibration due to a brighter limiting magnitude. In these cases, we used Landolt standard star fields \citep{landolt92,landolt09} to determine the zeropoint and airmass extinction coefficient.

Because of our limited observing window, asteroid 2008 HE was observed in crowded star fields and suffered significant contamination. Difference image analysis \citep[DIA;][]{bramich08,bramich13} was used to extract useful data from such images. The technique models the convolution kernel as a discrete pixel array, rather than a combination of linear functions. Each pixel value is solved for using a linear least squares. Using the kernel model, stars from a reference image can be blurred to match the seeing of each frame. After subtracting the blurred master from the image, the majority of the background stars are removed, allowing for more accurate photometry of the asteroid. This technique was only used for the images of 2008 HE because images of other asteroids are clean and the improvement form the usage of DIA is minimal. 

\subsection{Normalization}
We estimated the absolute magnitude $(H(1,1,0)$ (or $H$), which is the apparent $V$-band magnitude at 1 au from the Sun and the Earth, observed at a phase angle of zero degrees. For observations made using SDSS filters, the $g'$ magnitude value was converted to $V$ magnitude ($m_V$) according to \citet{jordi06}. To determine the absolute magnitude, we used the formula:
\begin{equation}
H(1,1,0)=m_V - 5\log_{10}(\Delta r_\mathrm{h}) + 2.5\log_{10}(\phi(\alpha))
\end{equation}
where $\Delta$ equals the geocentric distance in au, $r_\mathrm{h}$ equals the heliocentric distance in au, $\alpha$ is the phase angle (the Sun-Target-Observer angle) and $\phi(\alpha)$ is the phase integral, which is the ratio of the brightness at phase angle $\alpha$ to that at phase angle $0^{\circ}$. For the majority of the observed \lowq asteroids, we used $HG$ formalism \citep{bowell89} with $G = 0.15$ to determine the phase integral. Asteroid 2002 AJ129 was the only object with enough phase coverage to determine the phase integral using the three-parameter magnitude phase function \citep[$HG_1G_2$;][]{muinonen10}, which we then used to determine a more accurate $H$ (Section \ref{sec:phasecoef}). 
 
\subsection{Reflectance Values}
SDSS color filters provide sufficient wavelength coverage to study spectral slope trends \citep[e.g.,][]{thomas13, graves18, thomas21}. We followed techniques used by \citet{demeo13} and \citet{thomas13} to calculate the spectral slope over the $g'$ , $r'$ , and $i'$ reflectance values (hereafter $gri$-slope) to represent the slope of the continuum and $z'-i'$ color representing the depth of the possible 1 $\mu$m band. First, known Sloan filter solar colors\footnote{\url{http://classic.sdss.org/dr6/algorithms/sdssUBVRITransform.html}} ($g'-r'=0.44$, $g'-i'=0.55$, $g'-z'=0.58$~mag) were removed from the color indices. Then, the Sun-subtracted color indices were converted into reflectance values, normalized to the $g'$ band:
\begin{equation}
    \frac{R_x}{R_g} = 10^{0.4[(m_g-m_x)-(m_{g,\odot}-m_{x,\odot})]}
\end{equation}
The error for each reflectance point is calculated using standard error propagation. The $R_g$ value does not have an error as the other values are always in relation to $R_g = 1$. 

A linear regression was fit to the three photometric points using the central wavelength of the filters to calculate the slope. We computed the slope errors for each object via a Monte Carlo calculation, a technique used by \citet{thomas13,thomas21}. For each object, the individual reflectance values were modified by applying an offset of a random number pulled from a Gaussian distribution where the standard deviation is the 1-$\sigma$ error of the reflectance value. This calculation was done 20,000 times for each object and a slope was determined for each altered spectrum. The uncertainty of the slope was the standard deviation of the altered slopes generated by this process.

\begin{figure}[t!]
    \centering
    \includegraphics[width=0.5\textwidth]{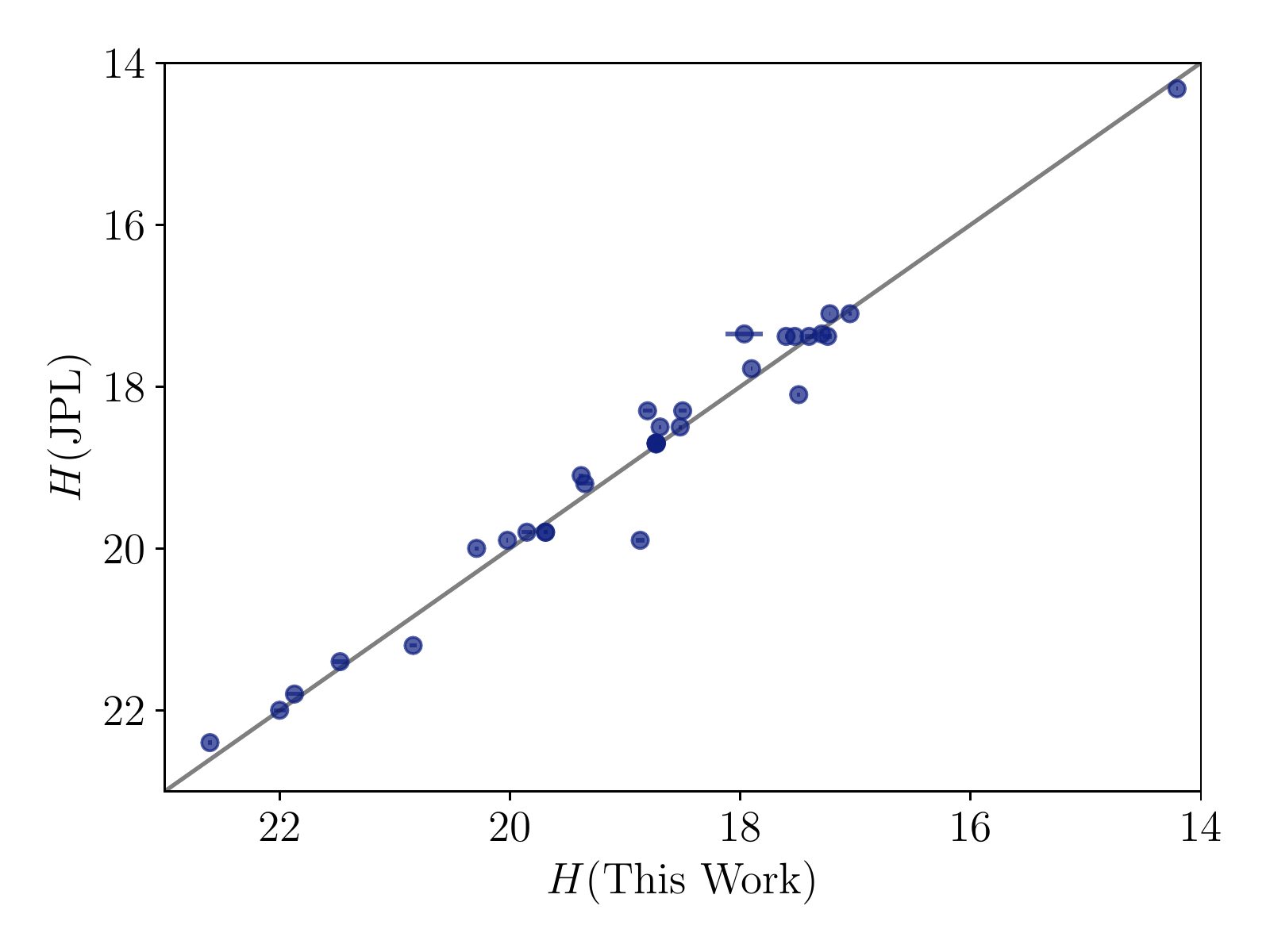}
    \caption{Absolute magnitude comparisons between this work and JPL. The solid line represents equality between the measurements. For all but one point, the errorbars are all within the marker. We find agreement between the values suggesting none of the objects were active during observations. All measurements including the largest difference between magnitudes of 1.02 (2008 MG1) are still within typical estimates for rotational variation. }
    \label{fig:Hmag}
\end{figure}

\section{Observational Results}
\label{sec:results}

\subsection{Absolute Magnitude H}
\label{sec:results:h}
We find our derived absolute magnitude values to be mostly consistent with those reported by the Jet Propulsion Laboratory's (JPL) Small-Body Database\footnote{\url{https://ssd.jpl.nasa.gov/}} with a standard deviation of the difference equal to $\sim$0.3 (Figure \ref{fig:Hmag}). An offset in absolute magnitude larger than the uncertainty due to rotational variability \citep[$A_{max}\lesssim$ 1~mag for small asteroids;][]{statler13} could indicate unresolved activity during one epoch. However, we do not observe any such offset and will explore upper limits of activity in Section \ref{sec:activity}.

\subsection{Color Distribution}

\begin{figure*}[t!]
    \centering
    \includegraphics[width=0.7\textwidth]{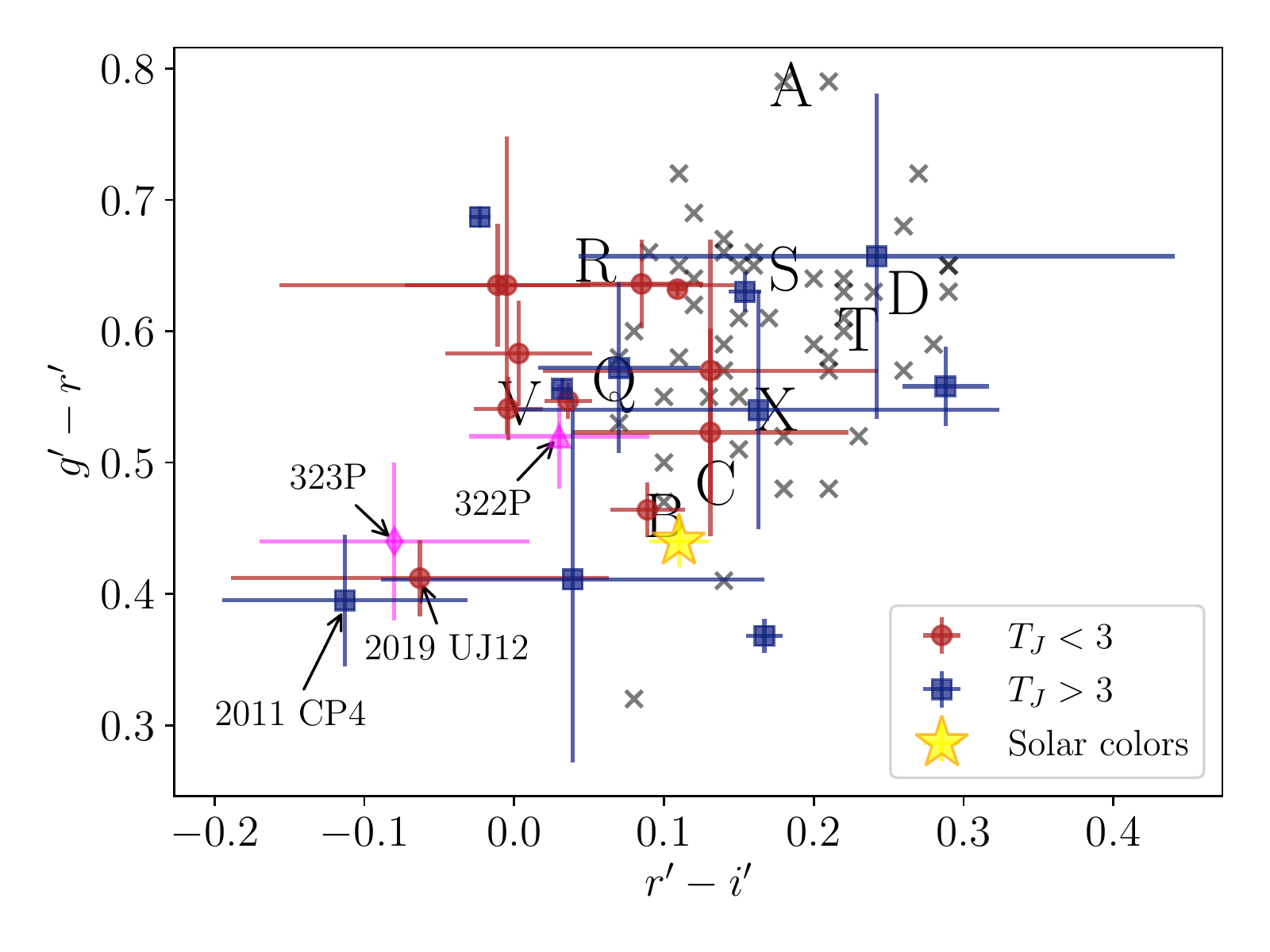}
    \caption{Color-color diagram of \lowq asteroids. The grey `x's are the color values of NEAs from the SDSS Moving Object Catalog \citep{ivezic01, ivezic02}. The large capital letters represent the average colors of taxonomic classes of NEAs from \citet{dandy03}. The color of comet 322P from \citet{knight16} is shown as a magenta triangle. The first reported color of 323P from \citet{hui22} is shown as the magenta diamond. The red circles and blue squares are \lowq asteroids with $T_J$ less than and greater than 3, respectively.
    }
    \label{fig:colorcolor}
\end{figure*}

Our measured colors are summarized in Table \ref{tab:obs} and plotted in Figure \ref{fig:colorcolor}. For near-Sun asteroids, we see a wide variety of colors with large overlap between the near-Sun distribution and the colors reported by \citet{dandy03} and SDSS. Some objects have error bars 
\clearpage
\begin{longrotatetable}
\begin{deluxetable*}{llcrrrcccccccrrr}
\tablenum{4}
\tablecaption{Summary of Observations and Measured Magnitudes\label{tab:obs}}
\tablecolumns{16}
\tablewidth{0pt} 
\tabletypesize{\notsotiny}
\setlength{\tabcolsep}{0.04in}
\tablehead{   
  \colhead{Object}&
  \colhead{UT Date}&
  \colhead{Tel.\textsuperscript{a}}&
  \colhead{$r_\mathrm{h}$\textsuperscript{b}}&
  \colhead{$\Delta$\textsuperscript{c}}&
  \colhead{$\alpha$\textsuperscript{d}}& 
  \colhead{Filters}&
  \colhead{$H_V$\textsuperscript{e}}&
  \colhead{$g'$}&
  \colhead{$r'$}&
  \colhead{$i'$}&
  \colhead{$z'$}&
  \colhead{$g'-r'$}&
  \colhead{$r'-i'$}&
  \colhead{$i'-z'$}&
  \colhead{$gri$-slope\textsuperscript{f}}
}
\decimals
\startdata
1995 CR&2017 Jan 25&INT&1.415&0.520&27.700&$g',r',i'$&21.87&$22.826\pm0.092$&$22.169\pm0.084$&$21.928\pm0.181$&-&$0.657\pm0.124$&$0.242\pm0.199$&-\tblspace&$1.36\pm1.02$\\
2000 BD19&2017 Jan 26&INT&1.177&0.606&56.700&$g',r',i',z'$&17.53&$19.260\pm0.008$&$18.554\pm0.003$&$18.551\pm0.008$&$19.120\pm0.034$&$0.706\pm0.008$&$0.003\pm0.008$&$-0.569\pm0.035$&$0.59\pm0.08$\\
&2020 Jan 28&31in&1.103&0.272&57.788&$B,V,R,I$&17.60&-&$17.071\pm0.007$&$16.654\pm0.007$&$16.494\pm0.009$&$0.547\pm0.023$&$-0.074\pm0.016$&-\tblspace&$-0.23\pm0.11$\\
&2020 Jan 30&31in&1.077&0.237&61.215&$B,V,R,I$&17.40&$17.624\pm0.057$&$16.621\pm0.037$&$16.265\pm0.035$&$16.317\pm0.054$&$0.447\pm0.092$&$-0.288\pm0.070$&-\tblspace&$-1.06\pm0.26$\\
&2020 Jan 31&31in&1.064&0.220&63.355&$B,V,R,I$&17.24&$17.378\pm0.043$&$16.337\pm0.038$&$15.888\pm0.033$&$16.119\pm0.034$&$0.600\pm0.090$&$-0.468\pm0.053$&-\tblspace&$-1.10\pm0.23$\\
2000 LK&2019 Jun 25&SOAR&2.421&1.839&22.766&$g',r',i'$&18.50&$23.223\pm0.051$&$22.555\pm0.037$&$22.516\pm0.065$&-&$0.668\pm0.063$&$0.040\pm0.075$&-\tblspace&$0.58\pm0.36$\\
&2020 Mar 1&LDT&2.091&1.373&23.257&$g',r',i',z'$&18.81&$22.548\pm0.048$&$21.955\pm0.052$&$22.073\pm0.096$&-&$0.592\pm0.071$&$-0.118\pm0.109$&-\tblspace&$-0.22\pm0.40$\\
2002 AJ129&2018 Feb 9&LDT&1.077&0.093&14.192&$g',r',i',z'$&18.73&$14.785\pm0.013$&$14.250\pm0.009$&$14.203\pm0.007$&$14.522\pm0.009$&$0.535\pm0.015$&$0.048\pm0.011$&$-0.319\pm0.012$&$0.12\pm0.09$\\
&2018 Feb 10&LDT&1.091&0.109&17.083&$g',r',i',z'$&18.73&$15.245\pm0.008$&$14.680\pm0.007$&$14.658\pm0.006$&$14.956\pm0.008$&$0.565\pm0.010$&$0.022\pm0.009$&$-0.298\pm0.010$&$0.14\pm0.08$\\
2002 PD43&2018 Jun 25&SOAR&1.239&0.836&54.575&$g',r',i',z'$&19.38&$21.795\pm0.029$&$21.211\pm0.028$&$21.209\pm0.040$&-&$0.583\pm0.040$&$0.003\pm0.049$&-\tblspace&$0.14\pm0.22$\\
2004 UL&2019 Jun 25&SOAR&1.631&1.793&34.124&$g',r',i',z'$&18.73&$22.727\pm0.115$&$22.316\pm0.077$&$22.277\pm0.102$&-&$0.411\pm0.139$&$0.039\pm0.128$&-\tblspace&$-0.31\pm0.56$\\
2006 HY51&2019 Mar 29&42in&2.585&2.643&22.387&$B,V,R,I$&17.05&$19.133\pm0.029$&$18.309\pm0.016$&$17.871\pm0.004$&$17.562\pm0.047$&$0.581\pm0.035$&$0.076\pm0.052$&-\tblspace&$0.38\pm0.23$\\
&2019 Mar 31&42in&1.470&0.660&34.530&$g',r',i',z'$&17.22&$18.960\pm0.004$&$18.326\pm0.004$&$18.218\pm0.005$&$18.578\pm0.007$&$0.634\pm0.006$&$0.109\pm0.006$&$-0.360\pm0.008$&$0.71\pm0.08$\\
2006 TC&2019 Oct 21&SOAR&1.446&0.878&42.540&$g',r',i',z'$&18.73&$21.636\pm0.143$&$20.902\pm0.029$&$21.113\pm0.056$&-&$0.523\pm0.153$&$0.211\pm0.064$&-\tblspace&$0.66\pm0.64$\\
2007 EP88&2017 Apr 7&SOAR&1.571&0.664&23.890&$g',r',i',z'$&18.70&$20.280\pm0.017$&$19.654\pm0.010$&$19.502\pm0.009$&$20.115\pm0.027$&$0.626\pm0.019$&$0.152\pm0.013$&$-0.613\pm0.028$&$0.85\pm0.12$\\
&2020 Jan 2&SOAR&0.962&0.663&72.030&$g',r',i',z'$&18.52&$20.442\pm0.019$&$19.805\pm0.016$&$19.643\pm0.016$&-&$0.637\pm0.025$&$0.162\pm0.023$&-\tblspace&$0.93\pm0.16$\\
2008 HE&2018 Mar 9&LDT&1.638&1.697&34.584&$g',r',i',z'$&17.49&$21.465\pm0.017$&$20.924\pm0.017$&$20.928\pm0.015$&$20.853\pm0.026$&$0.541\pm0.024$&$-0.004\pm0.023$&$0.075\pm0.030$&$-0.03\pm0.13$\\
2008 HW1&2017 Apr 7&SOAR&2.937&1.995&8.014&$g',r',i',z'$&17.97&$22.705\pm0.264$&$22.137\pm0.105$&$21.878\pm0.077$&-&$0.568\pm0.284$&$0.259\pm0.130$&-\tblspace&$1.04\pm1.31$\\
&2017 Jan 24&INT&2.194&1.802&26.249&$g',r',i',z'$&17.29&$21.845\pm0.031$&$21.208\pm0.014$&$21.142\pm0.040$&-&$0.637\pm0.034$&$0.066\pm0.043$&-\tblspace&$0.56\pm0.22$\\
2008 MG1&2017 Jul 1&SOAR&0.721&1.692&8.361&$g',r',i',z'$&18.87&$20.218\pm0.055$&$19.646\pm0.035$&$19.576\pm0.041$&$19.684\pm0.094$&$0.572\pm0.065$&$0.070\pm0.054$&$-0.108\pm0.103$&$0.33\pm0.31$\\
&2019 Jun 25&INT&1.371&0.373&15.476&$g',r',i',z'$&20.02&$19.745\pm0.009$&$19.197\pm0.006$&$19.177\pm0.014$&$19.260\pm0.017$&$0.547\pm0.010$&$0.020\pm0.015$&$-0.083\pm0.022$&$0.07\pm0.09$\\
2008 XM&2018 Jan 20&LDT&1.506&0.596&22.826&$g',r',i',z'$&20.29&$21.476\pm0.022$&$20.917\pm0.002$&$20.629\pm0.020$&$21.021\pm0.065$&$0.558\pm0.030$&$0.288\pm0.029$&$-0.392\pm0.068$&$1.11\pm0.19$\\
2010 JG87&2019 Aug 26&LDT&1.785&1.683&33.732&$g',r',i',z'$&19.35&$23.520\pm0.094$&$22.885\pm0.062$&$22.890\pm0.138$&-&$0.635\pm0.113$&$-0.005\pm0.152$&-\tblspace&$0.30\pm0.67$\\
2011 CP4&2017 Jan 24&INT&1.701&0.727&7.423&$g',r',i',z'$&20.84&$22.079\pm0.042$&$21.684\pm0.027$&$21.797\pm0.078$&-&$0.395\pm0.050$&$-0.113\pm0.082$&-\tblspace&$-0.77\pm0.25$\\
2011 KE&2018 Apr 14&LDT&1.755&0.800&14.953&$g',r',i',z'$&19.69&$21.528\pm0.020$&$21.085\pm0.022$&$20.957\pm0.028$&$21.444\pm0.072$&$0.443\pm0.029$&$0.128\pm0.035$&$-0.487\pm0.077$&$0.07\pm0.16$\\
&2018 Apr 15&LDT&1.768&0.820&15.624&$g',r',i',z'$&19.69&$21.645\pm0.021$&$21.157\pm0.022$&$21.111\pm0.030$&$21.512\pm0.082$&$0.488\pm0.031$&$0.046\pm0.037$&$-0.400\pm0.087$&$-0.04\pm0.17$\\
2013 YC&2017 Dec 17&LDT&1.431&0.447&0.856&$g',r',i',z'$&21.48&$20.986\pm0.051$&$20.416\pm0.086$&$20.284\pm0.072$&$20.544\pm0.054$&$0.570\pm0.100$&$0.131\pm0.112$&$-0.260\pm0.090$&$0.54\pm0.56$\\
2017 AF5&2017 Jan 24&INT&1.549&0.581&10.935&$g',r',i',z'$&17.90&$18.685\pm0.007$&$18.137\pm0.012$&$18.102\pm0.010$&$18.230\pm0.029$&$0.547\pm0.014$&$0.036\pm0.016$&$-0.128\pm0.030$&$0.12\pm0.10$\\
2018 GG5&2018 Apr 23&LDT&1.276&0.392&39.622&$g',r',i',z'$&19.86&$20.232\pm0.056$&$19.710\pm0.055$&$19.578\pm0.074$&$19.935\pm0.135$&$0.523\pm0.079$&$0.131\pm0.092$&$-0.356\pm0.154$&$0.36\pm0.44$\\
2019 AM13&2020 Mar 1&LDT&1.638&0.657&7.810&$g',r',i',z'$&22.00&$23.037\pm0.073$&$22.496\pm0.054$&$22.334\pm0.152$&$20.595\pm0.079$&$0.540\pm0.091$&$0.163\pm0.161$&-\tblspace&$0.54\pm0.68$\\
2019 UJ12&2019 Nov 4&LDT&1.145&0.192&34.023&$g',r',i',z'$&22.61&$20.984\pm0.012$&$20.572\pm0.026$&$20.635\pm0.123$&-&$0.412\pm0.029$&$-0.063\pm0.126$&$0.013\pm0.166$&$-0.60\pm0.35$\\
Phaethon&2017 Jan 25&INT&2.154&2.307&25.206&$g',r',i',z'$&14.21&$19.071\pm0.011$&$18.703\pm0.007$&$18.536\pm0.010$&$18.617\pm0.037$&$0.368\pm0.013$&$0.167\pm0.012$&$-0.081\pm0.039$&$-0.06\pm0.09$\\
\hline
2004 LG\textsuperscript{g}&2019 Mar 31&LDT&1.719&1.711&33.862&$g',r',i'$&17.67&$21.754\pm0.041$&$21.178\pm0.039$&$21.355\pm0.078$&-&$0.568\pm0.058$&$-0.012\pm0.100$&-\tblspace&-\tblspace\\
\enddata
\tablenotetext{\text{a}} {Telescope used: LDT = Lowell Discovery Telescope (4.3-m), 42in = Hall 42-in Telescope (1.1-m), 31in = 31-in Telescope (0.8-m), INT = Isaac Newton Telescope (2.5-m), SOAR = Southern Astrophysical Research Telescope (4.1-m)}
\vspace*{-3mm}
\tablenotetext{\text{b}} {Heliocentric distance, au}
\vspace*{-3mm}
\tablenotetext{\text{c}} {Geocentric distance, au}
\vspace*{-3mm}
\tablenotetext{\text{d}} {Phase angle, degree}
\vspace*{-3mm}
\tablenotetext{\text{e}} {Absolute $V$ magnitude measured by this work}
\vspace*{-3mm}
\tablenotetext{\text{f}} {The slope of the reflectance values over $g'$, $r'$, and $i'$ bandpasses [\%/$\mu m$]}
\vspace*{-3mm}
\tablenotetext{\text{g}} {2004 LG was formerly \lowq \citep{vokrouhlicky12b}}
\end{deluxetable*}
\end{longrotatetable}

\noindent spanning a broad spectrum, which we attribute to faint objects observed under less-than-ideal conditions after inspecting each frame individually to rule out uncertainty due to the occasional background star or poor photometric solutions.  In Fig.~\ref{fig:colorcolor}, we also separate objects by Tisserand parameter $T_J$. Most main-belt asteroids have $T_J > 3$, most Jupiter-family comets (JFCs) have $2 < T_J < 3$, and long-period comets (LPCs) have $T_J < 2$. Therefore, $T_J$, which is related to an object’s encounter velocity with Jupiter can generally be used to distinguish between types of orbits \citep{tisserand1896, kresak72, carusi87, levison96}. The red circles have a cometary orbit ($T_J < 3$) and the blue squares have an asteroidal orbit ($T_J > 3$). We do not see a clear trend with color versus $T_J$. The two objects with the smallest spectral slope in the lower left of the figure are 2011 CP4 and 2019 UJ12. There are several objects with $r'-i'$ colors most similar to V-types, but some of their $g'-r'$ colors are more red.

We compared the distribution of colors of \lowq objects to the range for NEAs using the Sloan Digital Sky Survey Moving Object Catalog \citep[SDSS, MOC, ][]{ivezic01, ivezic02}, which observed 471,569 moving objects through March 2007, using five filters, $u'$, $g'$, $r'$, $i'$, and $z'$. We restricted the sample from the SDSS MOC database according to the criteria outlined in \citet{demeo13} Section 2.1, which includes filtering faint data, data with large errors, and data with flags relevant to moving objects and good photometry. We further reduce the sample by only including objects that are categorized as NEAs ($q\leq1.3$~au). Applying the selection criteria, we are left with a sample of 42. We plot the range of colors as a gray `x's behind our \lowq measurements and the average colors of NEA types observed by \citet{dandy03} to give context for our findings. We also included the colors of 322P/SOHO 1 \citep{knight16}, 323P/SOHO \citep{hui22}, and solar colors.

In the color-color plot (Figure \ref{fig:colorcolor}), we observe that \lowq asteroids have a bluer distribution compared to average NEA colors (more in the lower left than the upper right). To investigate this further, we used a kernel density estimation (KDE) plot of the colors (Figure \ref{fig:kde}). Rather than using discrete bins, a KDE plot smooths the measurements with a Gaussian kernel, producing a continuous probability density estimate. The data points are weighted by their 1-$\sigma$ uncertainty and the kernel bandwidth is selected according to Scott's Rule \citep{scott92}. The result is shown in Fig.~\ref{fig:kde}. The \lowq $g' - r'$ colors appear to have the same distribution as the colors from SDSS MOC, but shifted bluer. The \lowq $r' - i'$ colors show a wider distribution than the SDSS MOC, but they extend to bluer colors. The KDE plots confirm our interpretation of the color-color plot.

\begin{figure*}[hbt!]
    \centering
    \includegraphics[width=0.9\textwidth]{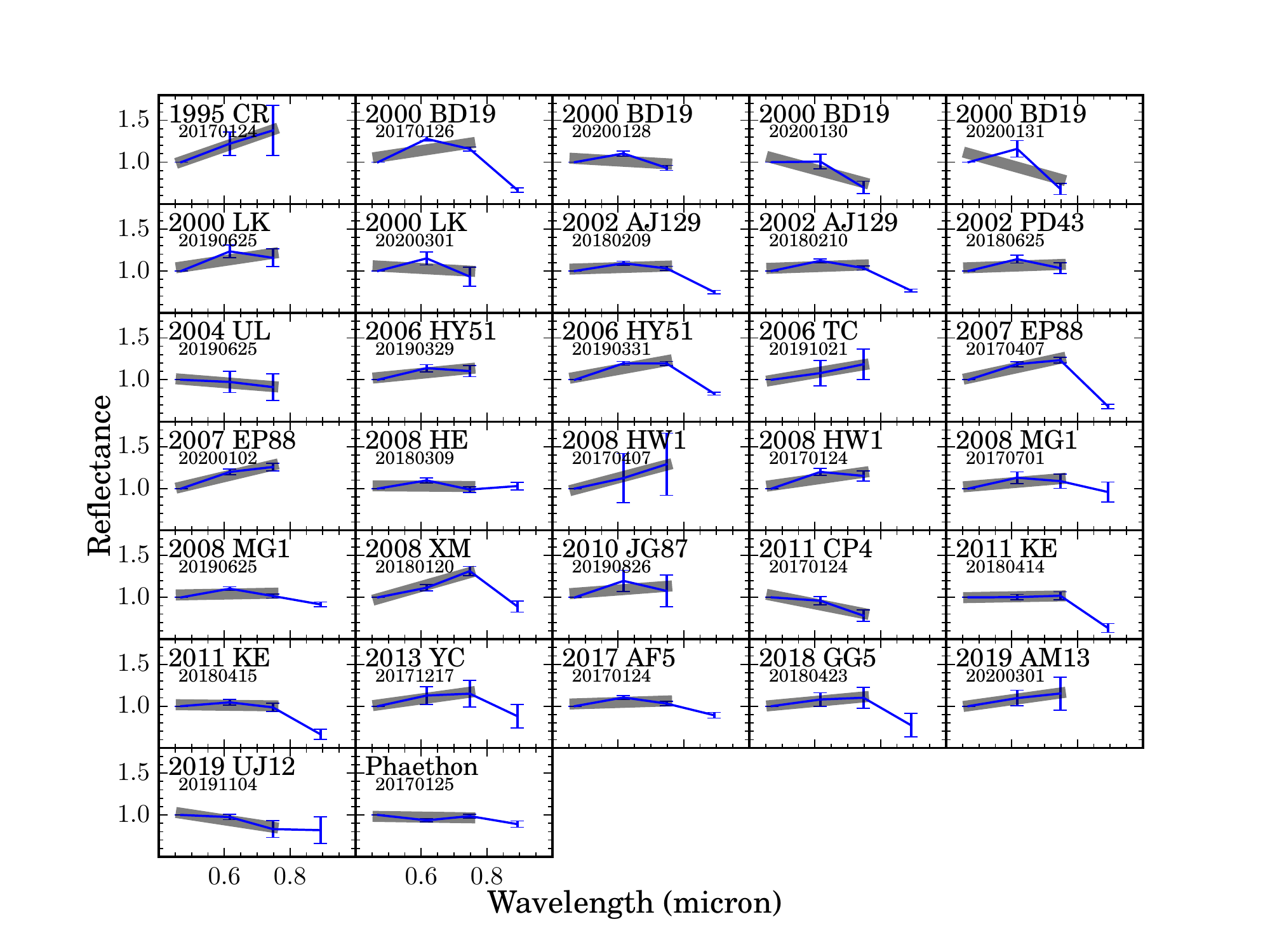}
    \caption{Spectrophotometry of \lowq asteroid observations created by transforming colors into reflectance values. Our measurements are in blue at their central wavelength, normalized to the $g'$ filter, with uncertainties shown. The grey bar shows the measured $gri$-slope. The object's name and date of observation (YYYYMMDD) are given on each panel.}
    \label{fig:spectra}
\end{figure*}

\begin{figure*}[hbt!]
    \centering
    \includegraphics[width=0.8\textwidth]{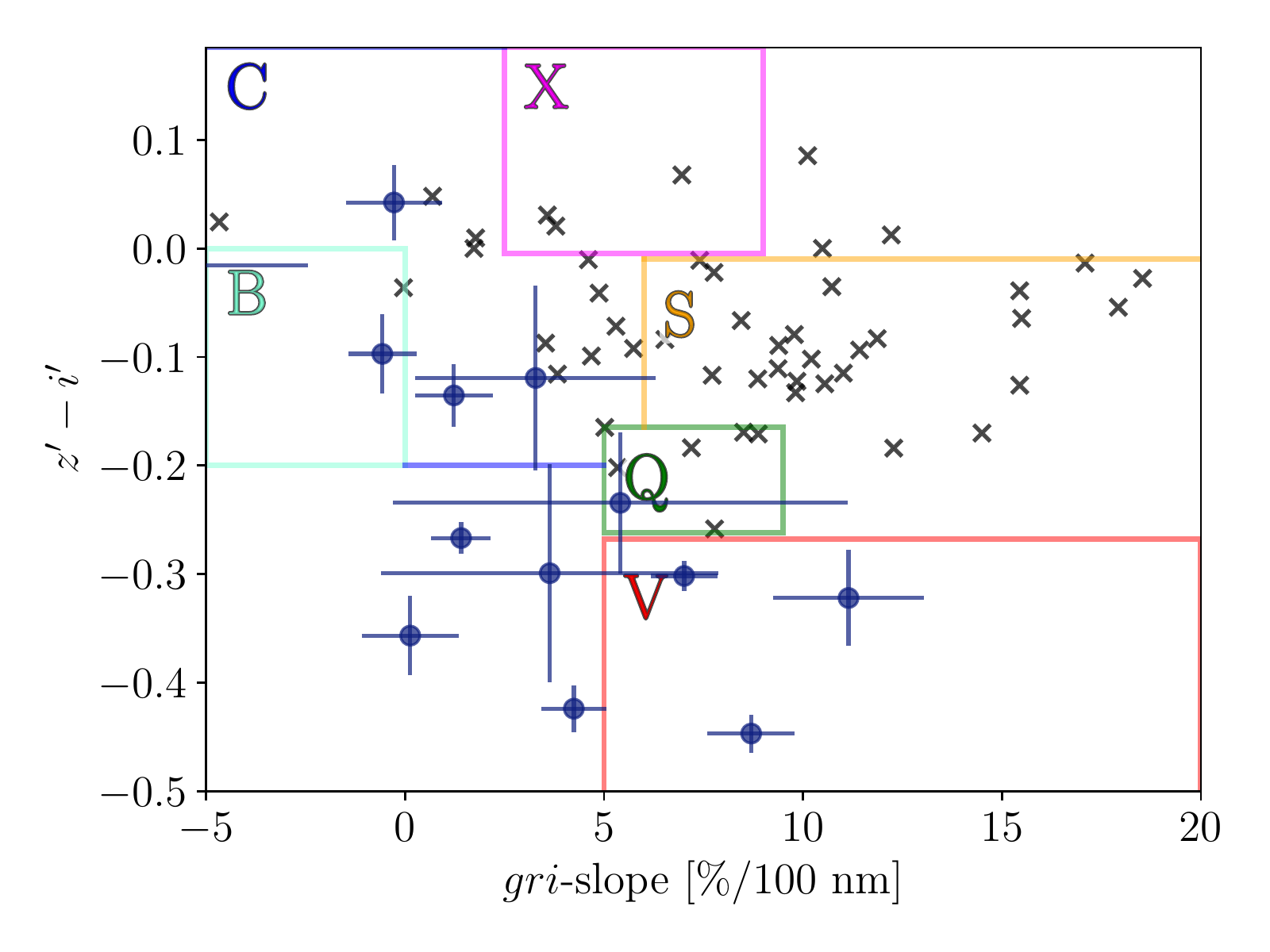}
    \caption{$gri$-slope and $z'-i'$ reflectance values of \lowq asteroids (blue dots). For comparison we show the distribution of NEAs from the SDSS MOC4 catalog (gray `x's) and boundaries used to classify SDSS data from \citet{demeo13}. Low-$q$ asteroids have a flatter spectral slope than the NEA population.}
    \label{fig:slope_iz}
\end{figure*}

The reflectance values for each observation can be seen in Figure \ref{fig:spectra}. We use the reflectance values to determine the $gri$-slope and $z'-i'$ color, representing the slope of the continuum and the 1-$\mu$m band depth, respectively. Both of these parameters are affected by space weathering, which causes a steeper spectral slope and a shallower 1-$\mu$m band depth. We compare our sample to NEAs from SDSS MOC values and class boundaries reported by \citet{demeo13} (Figure \ref{fig:slope_iz}). We find near-Sun asteroids have a shallower spectral slope compared to the NEA population, especially those with V-type colors, which agrees with our findings of bluer colors shown in Figure \ref{fig:colorcolor}. The $z'-i'$ reflectance values are smaller as well, consistent with a steeper 1-$\mu$m band depth.

\subsection{Spectral Slope Trends with Decreasing Perihelion}

\begin{figure}[b!]
    \centering
    \includegraphics[width=0.5\textwidth]{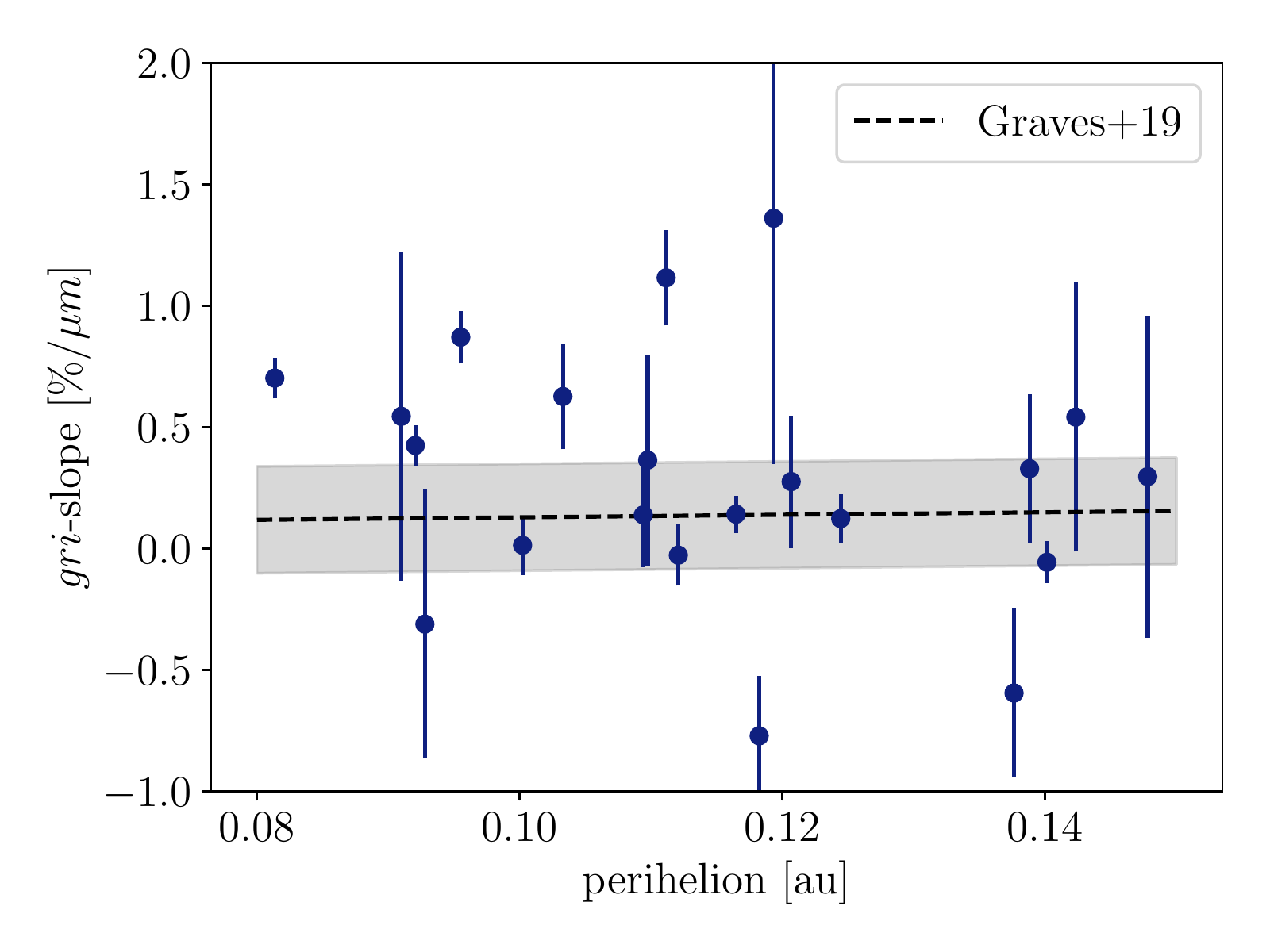}
    \caption{The distribution of $gri$-slope vs. perihelion distance of \lowq asteroids. The dashed line and the shaded region show the trend for S- and Q-type NEAs using data in \citet{binzel04,lazzarin04,lazzarin05} and the uncertainty at a 95\% confidence level according to \citet{graves19}. Our data are consistent with \cite{graves19} at the 1-$\sigma$ level.
    \vspace{5mm}}
    \label{fig:slope_q}
\end{figure}

 We compare the \lowq spectral slopes with the trend versus perihelion distance seen in S- and Q-type NEAs with $q\geq0.2$ au by \citet{marchi06a} and \citet{graves19} to determine if the same trend might extend to smaller perihelion distances (Figure \ref{fig:slope_q}). We can make this comparison because we anticipate our sample to be predominately S- or Q-type because together they are the most common NEA types \cite{binzel19} and based on dynamical modeling of source regions, which we discuss in Section \ref{sec:sourceregions}. For our comparison, we used the trend measured by \citet{graves19} who implemented a windowed moving average, instead of the similar trend determined by \citet{marchi06b}, who implemented a point-based moving average. Our data are consistent with the extended trend at the 1-$\sigma$ level. However, given the large uncertainty in some of the slopes in our dataset and the short range of $q$, the trend is virtually indistinguishable from a flat distribution, indicative of no spectral slope change with perihelion distance. Further investigation is warranted.

\subsection{Lightcurves}
We compiled high-cadence lightcurves of three bright objects (394130 (2006 HY51), 276033 (2002 AJ129) and 137924 (2000 BD19)) in our sample in order to measure their rotation periods and amplitudes. The methodology for determining the rotational period follows our approach in earlier papers \citep[e.g.,][]{knight11a, knight12, eisner17}. Because observing geometry changed rapidly throughout the night, we corrected our photometric results for the geometric circumstances image by image. The rotation period was estimated by superimposing the lightcurves from all nights, with the data phased to a ``trial" period and zero phase at perihelion. We then iterated the trial period, making direct ``better or worse" comparisons between each iteration, until a rotation period estimation was constrained by eye. We estimate the uncertainty by determining how much the period can be adjusted before a phased light curve appears obviously incorrect.

\subsubsection{2006 HY51}
\begin{figure}[H]
    \centering
    \includegraphics[width=0.42\textwidth]{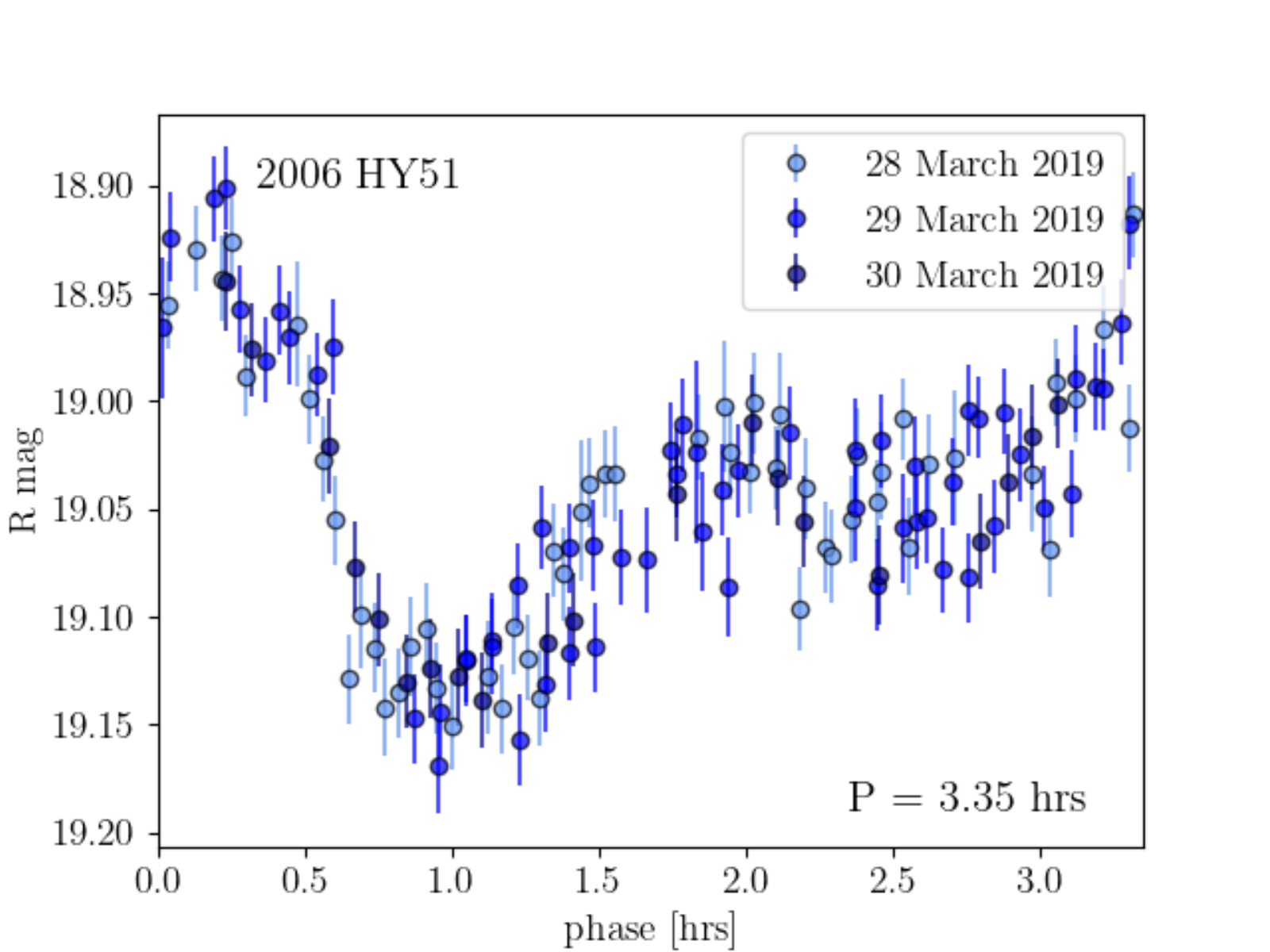}
    \caption{Rotational lightcurve of 2006 HY51 using data from March 2019 using Lowell Observatory's 42-inch telescope. The lightcurve is phased to the best-fit period of $3.350 \pm 0.008$ hours.}
    \label{fig:2006HY51rot}
\end{figure}

2006 HY51, whose rotation period was previously unconstrained, was observed over three nights (March 27-29, 2019) using Lowell Observatory's 42-inch telescope and the Johnson $R$ filter. We collected $\sim6$ hours of data each night with frames every $\sim300$ seconds. We measured a rotation period of $3.350 \pm 0.008$ hours and a maximum peak-to-trough amplitude of $\sim$0.2 mag (Fig. \ref{fig:2006HY51rot}). The lightcurve has an interesting shape, with one peak larger than the other, suggesting the shape of the asteroid deviates substantially from a tri-axial ellipsoid \citep{magnusson86}.

\subsubsection{2002 AJ129}
2002 AJ129 was observed over two nights (February 6-7, 2018) using Lowell Observatory's 31-inch telescope in robotic mode and the Johnson $R$ filter, collecting $\sim9$ hours of data each night with frames every $\sim90$ seconds. 
\begin{figure}[H]
    \centering
    \includegraphics[width=0.5\textwidth]{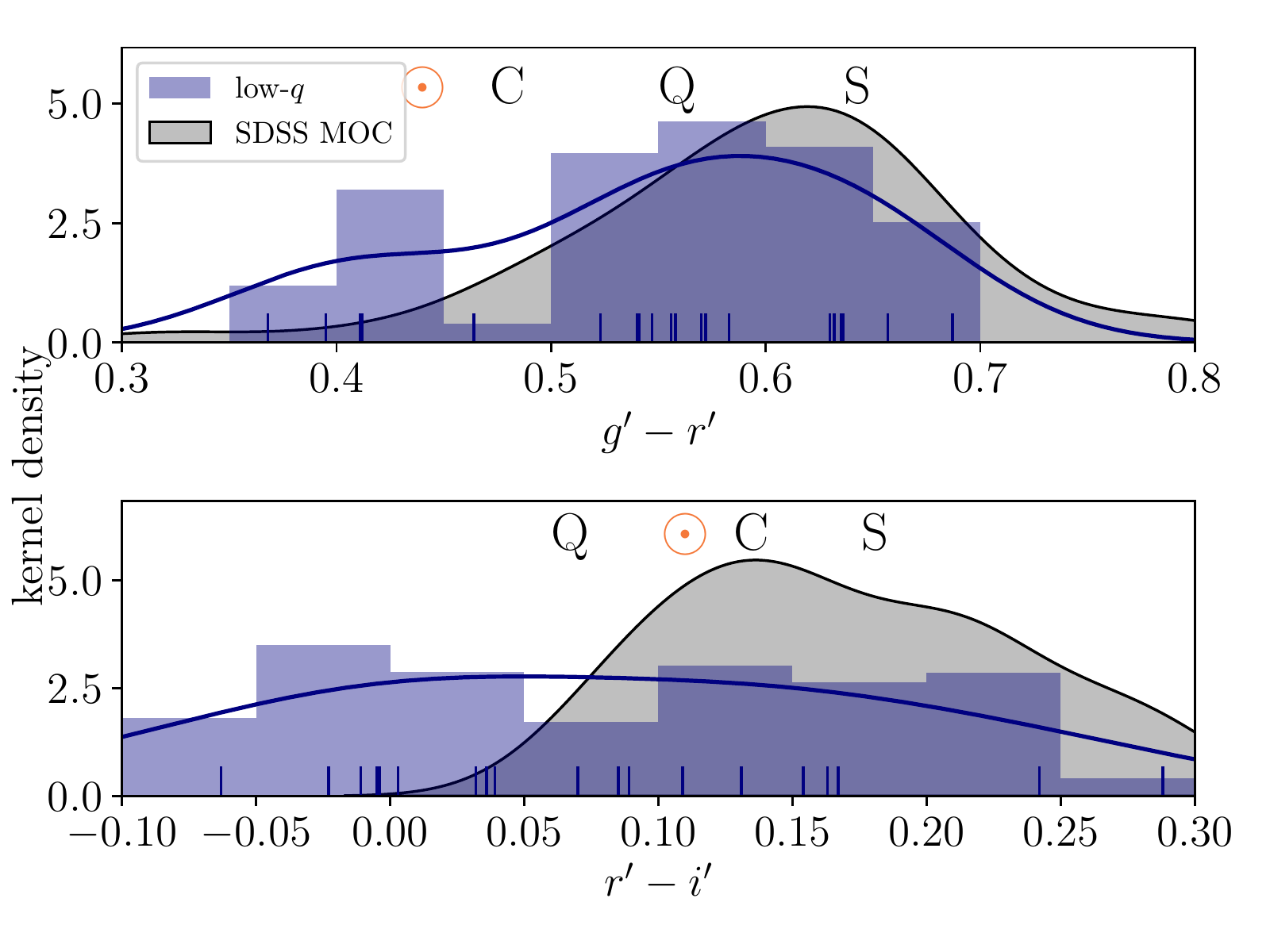}
    \caption{Kernel density plot of \lowq asteroid colors with underlying histogram (blue), where data points are weighted by their 1-$\sigma$ uncertainty and the kernel bandwidth is selected according to Scott's Rule \citep{scott92}. This allows us to assess the color distribution as a smooth probability density, instead of only discrete bins in the histogram. The black filled in curve is the color distribution of NEAs from SDSS MOC. The colors of the most common asteroid types according to \citet{dandy03} are shown as letters, as well as Solar colors shown as an orange ``$\odot$''. We find a preference towards bluer colors for both $g'-r'$ (top) and $r'-i'$ (bottom).}
    \label{fig:kde}
\end{figure}

\begin{figure}[H]
    \centering
    \includegraphics[width=0.42\textwidth]{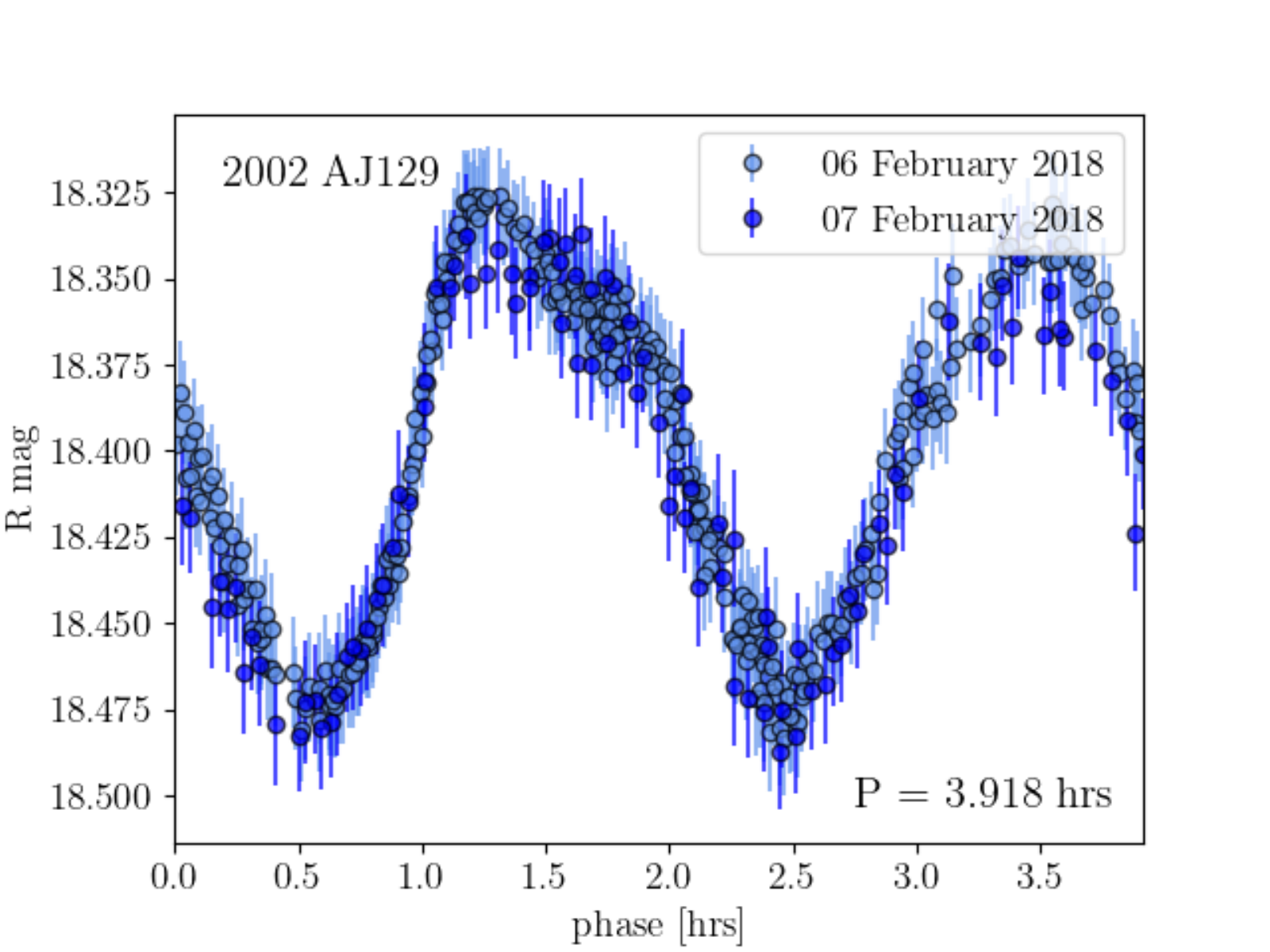}
    \caption{Rotational lightcurve of 2002 AJ129 using data from February 2018 using Lowell Observatory's 31-inch telescope. The lightcurve is phased to the best-fit period of 3.918 hours.}
    \label{fig:2002AJ129rot}
\end{figure}

\noindent We found a rotational period of $3.918 \pm 0.010$ hours and a peak-to-trough amplitude of $\sim$0.15 mag (Figure \ref{fig:2002AJ129rot}). The prepublished period of 2002 AJ129 was reported by the Ondrejov Asteroid Photometry Project \citep{pravec18}\footnote{\url{http://www.asu.cas.cz/\~ppravec/newres.htm}} 
as $3.9226 \pm 0.0007$, and \citet{devyatkin22} reported a period of $3.9222 \pm 0.0008$. Both measurements are within our uncertainty.

\subsubsection{2000 BD19}
\begin{figure}[H]
    \centering
    \includegraphics[width=0.42\textwidth]{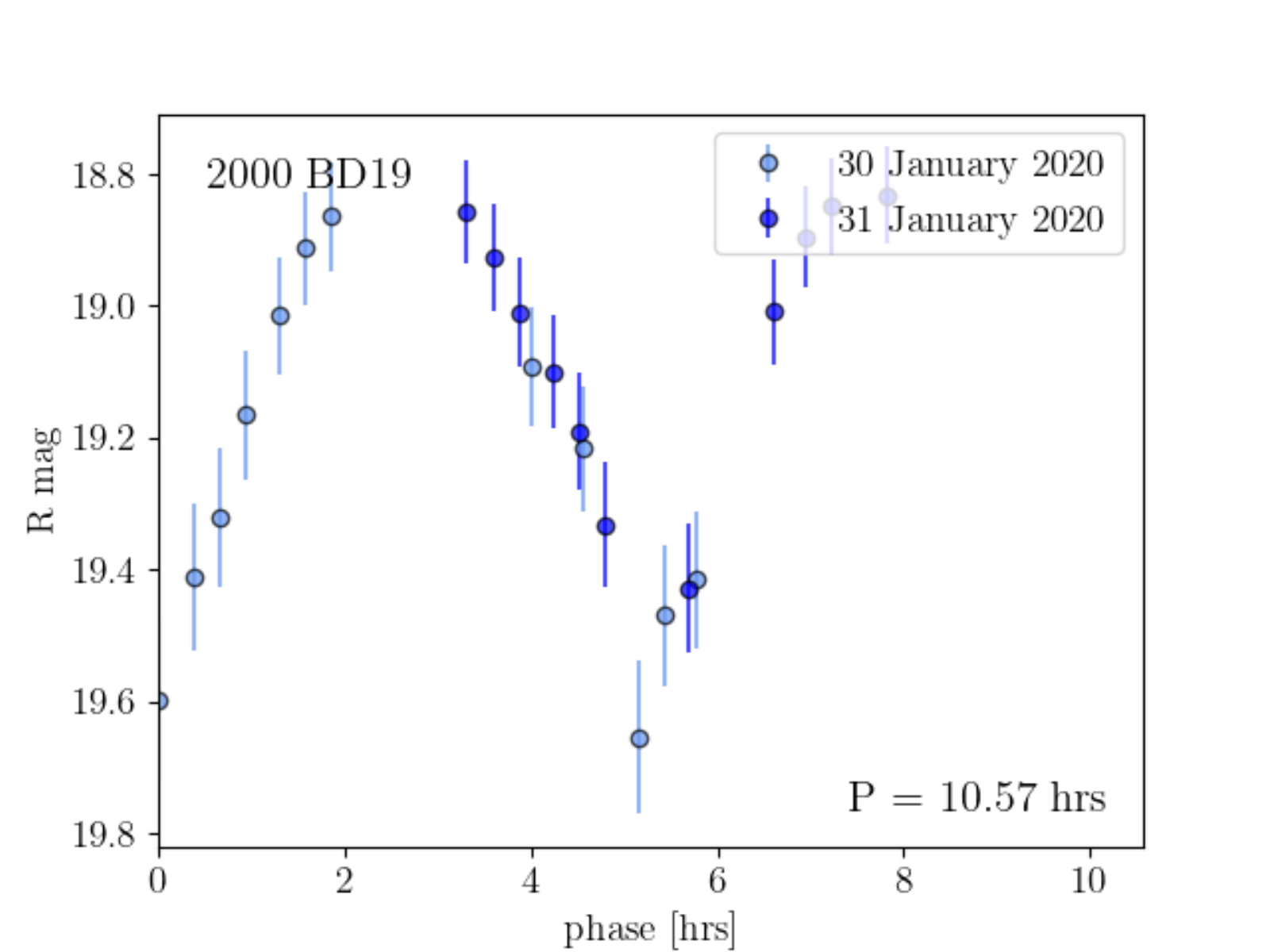}
    \caption{Rotational lightcurve of 2000 BD19 using data from January 2020 using Lowell Observatory's 31-inch telescope. The lightcurve is phased to the best-fit period of 10.57 hours determined by \citet{mpb2015july}. We were unable to constrain the rotational period further.}
    \label{fig:2000BD19rot}
\end{figure}

We observed 2000 BD19 over two half-nights (January 30--31, 2021) using Lowell Observatory's 31-inch in robotic mode and the Johnson $R$ filter, collecting $\sim6$ hours of data each night with frames every $\sim20$ minutes. \citet{mpb2015july} observed 2000 BD19 over six nights and report a rotational period of $10.570 \pm 0.005$ hours and a max peak-to-trough amplitude of $0.69\pm0.04$. Without full coverage, we are unable to constrain the rotational period further. However, when phased to the published period, our results are consistent with the reported period and amplitude (Figure \ref{fig:2000BD19rot}).

\subsection{Phase Coefficient}
\label{sec:phasecoef}
For most of the objects we observed, the range of phase angles was too small to determine the asteroid's phase function. The exception is asteroid 2002 AJ129, which we observed over two nights with Lowell Observatory's 31-inch over $\sim0.8$ -- $17.8$ degrees. Using the online implementation of the model selection stated in \citet{pentilla15}, we employ the $H,G_1,G_2$ system described in \citet{muinonen10}. G$_1$ and G$_2$ are determined to be $0.157\pm0.035$ and $0.486\pm0.017$, respectively, which is consistent with measurements of main-belt asteroids  \citep{muinonen10,veres15}.

\subsection{Source Region Probabilities}
\label{sec:sourceregions}

We assessed the origin of \lowq asteroids using the \citet{granvik18a} escape region model, which has been progressively developed by \citet{granvik16,granvik17,granvik18b}. Given the orbital elements and $H$ magnitude of an object, this model gives the probabilities that an object ``escaped'' from each of the seven source regions, where the sum of the probabilities is equal to one. The source regions consist of the $\nu_6$ inner main-belt region, Jupiter resonance complexes: 3:1, 5:2 and 2:1, the Hungarias, the Phocaeas, and the Jupiter-family comets. Source regions and orbital elements are tabulated for all 53 known \lowq objects in Table \ref{tab:orbits} and the source region with the highest probability of being the escape region for each object is in bold. All known objects with $q\leq0.15$ au most probably escaped from $\nu_6$ or 3:1 regions with the exception of 2010 JG87, which most likely escaped from the 5:2 region.

This is unsurprising considering the $\nu_6$ region and the 3:1 are the resonance escape regions for nearly 80 percent of NEAs \citep{bottke02, granvik18b, binzel19}. NEAs from these source regions are mostly S- or Q- type, with $\sim80-90$ percent having a geometric albedo greater than 0.1 \citep{binzel19, morbidelli20}. These resonances call for large oscillations of the eccentricity, so even if asteroids from these regions are not currently low-$q$, they possibly were at one point in their dynamical history. Around 80 per-
\clearpage

\vspace{-25pt}

\vspace{-25pt}

\noindent cent of all NEAs have experienced a perihelion distance smaller than 0.15 au at some point since they escaped from the main belt \citep{toliou21}. For our specific sample, we use the lookup table from \citet{toliou21}, which utilizes the NEO model from \citet{granvik18b} to determine the probability an object with a given set of orbital elements and $H$ magnitude spent time with a perihelion distance smaller than a certain $q_s$ and the estimated dwell times with $q\leq q_s$. Unlike previous models, the \citet{granvik18b} model accounts for ``supercatastrophic'' disruption of NEAs within a certain $q$, dependent on $H$. We tabulate the estimated dwell times each of our objects spent with $q\leq0.15$ au in Table \ref{tab:orbits}. Dwell times range from one kyr (2013 YC) to 15 Myrs (2007 EP88), consistent with the findings from \citet{marchi09}. The time spent with $q\leq0.15$ au can vary from several hundred years to a few Myr, averaging $\sim20$ kyr for NEAs from $\nu_6$ \citep{toliou21}. Asteroids escaped from the 3:1 resonance and outer main-belt spend an average of $\sim5$ kyr and $\sim300$ yrs with $q\leq0.15$ au \citep{marchi09}.

\section{DISCUSSION}
\label{sec:discussion}

\subsection{Near-Sun Processes}
Space weathering is the alteration of asteroid surfaces due to the space environment, which is dominated by ion radiation from the solar wind \citep{marchi06a,vernazza09,brunetto15}, as opposed to micrometeorite impacts, which have longer timescales \citep{sasaki01}. How space weathering affects an object is dependent on the object's original composition \citep{lantz17}. Because characterized \lowq objects typically have high albedos, where more than ten percent of incident radiation is reflected \citep{mainzer12, granvik16}, we focus on the space weathering effects on high-albedo objects. The space weathering effects on silicates are well understood, where nanophase reduced iron particles (npFe0) darken and spectrally redden the surface \citep[e.g.,][]{pieters00,taylor01,hapke01,clark02, brunetto15}. Space weathering of silicates will also suppress the pyroxene-olivine absorption bands \citep[e.g.,][]{pieters00, hapke01, gaffey10}. Therefore, when a fresh unweathered surface is exposed by some resurfacing process, the spectral slope becomes bluer for asteroids with a higher albedo. This has been observed on NEA Eros by NEAR \citep{clark01} and on regolith grains brought from NEA Itokawa during the Hayabusa mission \citep{noguchi11}, both of which are S-type asteroids.

There are several potential resurfacing mechanisms that affect asteroids when they closely approach the Sun. (a) Increasing temperatures can cause material from lower surface layers to sublimate and allow for progressively less volatile material to reach sublimation temperatures. Decomposition and sublimation of refractory organics begins around $\sim$450 K, metal sulfides at $\sim$700 K, and silicates at 1000--1500 K \citep[see review by][]{jones17}. (b) Many studies have explored resurfacing caused by the tidal forces experienced during planetary encounters \citep[e.g.,][]{nesvorny05, marchi06a,binzel10,nesvorny10,demeo14, carry16,devogele19}. (c) An asteroid with an irregular surface can experience an increase in spin rate due to asymmetrical radiative torques, known as the Yarkovsky-O'Keefe-Radzievskii-Paddack (YORP) effect \citep{bottke06}, which may result in centrifugal loss of material \citep{rubincam00,vokrouhlicky15}. (d) Thermal fatigue induced by the diurnal temperature variations can cause boulders and grains to fracture and break down into smaller regolith, which can then be lost due to outgassing and/or solar radiation pressure \citep{jewitt12,delbo14}. Such thermal cracking was observed on asteroid Bennu during the OSIRIS-REx mission \citep{molaro20}. (e) Resurfacing through impacts by high-speed near-Sun meteoroids that could eventually lead to disruption has also been suggested \citep{wiegert20}. All of these processes can resurface weathered regolith on \lowq asteroids, which would most likely cause an overall bluer spectrum.

\subsection{Colors} 
We found evidence for bluer colors in the \lowq population when compared to NEAs \citep{dandy03}, but it is not as obvious as we might expect given the bluer colors observed in comets 322P, 323P, and 96P \citep{knight16,eisner19}. Additionally, we find agreement with the extended bluening trend with decreasing perihelion seen in S- and Q-type NEAs at $q>0.2$ au \citep{marchi06b,graves19}. However, uncertainties are such that over the short range of $q$, the trend is indistinguishable from no spectral slope change with $q$.  

The color distribution of \lowq asteroids may be more stochastic because of competing processes and the potential variety in asteroid properties, including dynamical properties and regolith properties, such as albedo and grain size, which play a role in the extent of heating and near-Sun processing effects. Phase reddening, where the spectral slope increases with increasing solar phase angle, $\alpha$, is expected to contribute to the scatter in the distribution; the effect is common among S-group asteroids \citep{sanchez12, carvano15, perna18}, but the extent varies for each individual asteroid and therefore cannot be modeled out of our measurements \citep{carvano15,binzel19,popescu19}. 

\subsubsection{Dynamical Considerations}
\label{sec:disc_dynamics}

Near-Sun processes are generally influenced by the surface temperature and the length of time spent at that temperature. Therefore, understanding the dynamical history of \lowq objects is important for interpreting our results. While dynamical modeling can give a general idea, we cannot know for certain the dynamical history of any individual object. Some objects could have recently migrated to a \lowq orbit in the last couple hundred years, while others could have been in a stable \lowq orbit for several thousand years and some objects might have had \lowq in the past and have migrated back for the $n$-th time through resonance oscillations \citep[i.e., Lidov-Kozai oscillations:][]{lidov62, kozai62}. There is likely a broad range of dynamical histories for our sample. However, we can use dynamical modeling to better understand the temperatures experienced or the duration of heating throughout an individual asteroid's lifetime.

Roughly half of the known near-Sun asteroid population have $T_J<3$, which indicates a cometary orbit. If such asteroids are actually dormant comets, we might expect their colors to be distinct due to compositional differences or surface changes due to sublimation of volatiles \citep{jewitt15}. The percentage of NEAs that are dormant comets is estimated to be anywhere from 3 -- 20 percent \citep[e.g.,][]{fernandez05,bottke02,demeo08, mommert15}. We see no evidence for color difference based on Tisserand parameter with respect to Jupiter. We find the average colors for objects with asteroidal orbits to be $g'-r'= 0.54 \pm0.11$ and $r'-i'= 0.10\pm0.12$ and cometary orbits to be $g'-r'= 0.56 \pm 0.07$ and $r'-i'= 0.05 \pm 0.06$. However, we find that the colors are significantly bluer than cometary nuclei ($g'-r'= 0.68\pm0.06$ and $r'-i'= 0.23\pm0.04$; \citealt{jewitt15}). Because dormant comets make up $\sim10$\% of the NEA population, a larger sample size is needed to investigate such a connection. 

Low-$q$ asteroids 2019 UJ12 and 2011 CP4 have the bluest colors in our sample. The \citet{toliou21} model shows an average dwell time of 200 yrs for the orbital properties of 2019 UJ12, suggesting it is likely experiencing near-Sun resurfacing processes, such as thermal fatigue, for the first time. However, 2019 UJ12 has a poorly constrained orbit (U = 8), so we consider this result inconclusive. Other than 2019 UJ12, we observe no clear trend between dwell times and color. Although, it is important to note that the \citet{toliou21} model uses an integration over the entire dynamical lifetime of an object. To compliment the \citet{toliou21} model, we conducted backward-integration of the orbits of individual \lowq objects over the last 2000 yrs using the methods described in \citet{hsieh21} and found that 2011 CP4 has had several close encounters with Earth in the past 2000 yrs, which suggests its very blue color may be due to resurfacing caused by tidal encounters. However, space weathering occurs on timescales of $\sim10^4 - 10^6$ yrs, so potential resurfacing events occurring more than 2000 yrs ago must be considered, but such integrations are beyond the scope of this project. 

Formerly \lowq asteroid 2004 LG, which spent 2500 years with $q<0.076$ au \citep{vokrouhlicky12b,wiegert20} does not exhibit any unusual colors compared to the near-Sun population and other NEAs, though we would suspect otherwise after being in an extreme environment, within the catastrophic disruption limit determined by \citet{granvik16}. 2004 LG may not be an S- or Q-type and therefore, experiences near-Sun properties differently. There is also the possibility that surface variability over time may play a role, which will be discussed below.

\subsubsection{Varying Surface Properties}

Albedo, emissivity, macroscopic surface roughness, and thermal inertia all affect surface temperature. Surface roughness alone can raise the average surface temperature by 20--30\% \citep{marchi09}. Albedo affects the degree of space weathering and the amount of radiation absorbed vs.\ reflected. And thermal inertia governs the temperature distribution on an asteroid's surface. These surface properties, as well as rotational properties, influence the degree of thermal fatigue, which is thought to be the dominant regolith production process and resurfacing mechanism for near-Sun asteroids \citep{delbo14,graves19} and involves the formation and expansion of cracks as a result of diurnal temperature variation. These properties vary with composition, but can also vary between objects of the same taxonomy. Because most of our objects have few observations, these properties are essentially unknown but are expected to vary between objects, contributing to the observed scatter in the color measurements. 

There are more \lowq asteroids with colors similar to V-types than expected given that V-types make up only $\sim$5\% of the NEA population \citep{binzel19}. This could simply be small number statistics. However, V-type asteroids have high albedos ($\sim$0.3; \citealt{usui13}), so the overabundance could be due to an observation bias. Alternatively, as proposed by \citet{granvik16}, higher albedo objects are less likely to experience enough thermal cracking or sublimation to disrupt at \lowq because they absorb less heat than lower albedo objects. Thus, there may be a survivorship bias that favors V-type asteroids. Another possibility is that Yarkovsky drift may create a slower orbital progression timescale for higher albedo objects. If so, a longer residence lifetime in near-Sun orbits could create a bias toward these objects being noticed in a survey. However, \citet{granvik18b} found Yarkovsky drift on the NEO steady state orbit distribution to be negligible compared to the gravitational perturbations caused by planetary encounters. Thus, such an effect may be difficult to detect.

\subsubsection{Competing Processes}

Competing processes are likely contributing to the wide color distribution we measured compared to the NEA population as shown in the $r'-i'$ KDE plot in Fig. \ref{fig:kde}. The space weathering rate increases by two orders of magnitude from 1 to 0.1 au because solar irradiation increases as $r_\mathrm{h}$ decreases \citep{marchi06a,paolicchi07}. Similarly, several resurfacing processes are expected to increase at low-$q$. The effects from Yarkovsky and YORP are expected to increase with solar irradiation as well. As surface temperature increases, the diurnal temperature variation increases, allowing for more resurfacing due to thermal fatigue. Space weathering caused by solar irradiation has a timescale of $\sim10$ kyr -- 1 Myr \citep{hapke01,brunetto05,strazzulla05,brunetto06,loeffler09,vernazza09}. The timescale of thermal degradation is thought to decrease with decreasing heliocentric distance as $r_\mathrm{h}^k$ where $k > 5$ and the timescale for space weathering should be less than half the timescale of thermal degradation to explain the decreasing spectral slope with decreasing perihelion distance trend \citep{graves19}. However, those timescales were estimated based on S- and Q-type NEAs with $q>0.2$ au; those relations might not be valid at $q\leq0.15$ au. Additionally, multiple resurfacing processes can be occurring on the same object at the same time, which can further counter space weathering.

It is possible that the wide variety of colors we measured in the \lowq population is related to the consistent battling of processes. The refreshed surface continues to be irradiated by the Sun as it is undergoing resurfacing processes. As time in \lowq orbit passes, there is less and less ``fresh'' surface to expose. In other words, the resurfaced regolith has already been irradiated. \citet{binzel19} suggest that after enough time, the asteroid surface may be ``saturated'' by space weathering and will have maximum redness, while objects new to \lowq might have the most refreshed surface. The result would be a larger variance in colors with decreasing distance to the Sun. Comparing the \lowq sample with more distant NEAs, we find that the variance in $r'-i'$ color (0.010) is higher for our \lowq sample compared to the SDSS MOC NEA population (0.004), but the variance in $g'-r'$ color is the same for both (0.008). This theory could also explain the bluer colors of 2019 UJ12, as well as the redder color of 2007 EP88, which has spent the most time with $q\leq0.15$ au.

\subsection{Activity}
\label{sec:activity}
This study focused on observing surface properties since the goal was to sample the whole population in $\sim3$ years. As a result, most were too faint to meaningfully search for comae. We visually inspected all data and did not detect any evidence of activity, e.g., tail, coma, or broadened PSF. Comparing absolute magnitudes at the time of observation to previous data, we can crudely constrain that the \lowq asteroids had no activity in excess of typical NEA rotational amplitudes. 

Since all objects in our sample reach perihelion within SOHO's FOV, they are observable every orbit when they are most likely to be active. SOHO's limiting magnitude of $\sim$8 is not sensitive enough to detect any objects in our sample unless they are active. However, because forward-scattering of comet dust grains is highly efficient \citep[e.g.,][]{marcus07b, hui13}, for some orientations only a small amount of coma is needed in order to be detected. We constructed a simple model to estimate the observability of the \lowq asteroids when in the SOHO FOV  ($r_\mathrm{h}\leq0.15$\,au). Using the limiting magnitude of SOHO, we can determine the dust cross-section needed to be detectable:
\begin{equation}
C=\frac{(2.25 \times 10^{22}) \pi r_\mathrm{h}^2 \Delta^2}{p_r\phi(\alpha)}10^{0.4(m_{\odot}-m_{lim})}
\end{equation}
where $p_r$ is the $r'$-band geometric albedo, $\alpha$ is the phase angle (the Sun-target-observer angle), $\phi(\alpha)$ is the phase function \citep[we use the Schleicher-Marcus comet dust phase function;][]{schleicher11}, $m_{\odot}$ is the apparent $r'$-band magnitude of the Sun, and $m_{lim}$ is the limiting magnitude of SOHO ($\sim$8). We identified all instances when each asteroid in our sample was within the SOHO FOV. We then estimated the depth of global material lost in order for the object to be detectable by SOHO, assuming an average particle size of 1~$\mu$m, a density of 3000~kg/m$^3$,  an albedo of 0.15, and the diameter estimate for each object. The minimum depth was $\sim$0.2 ${\mu}m$, though most required more than 1~${\mu}$m, the approximate depth of material needed to be excavated by Phaethon to produce the amount of dust observed while Phaethon was active near perihelion \citep{li13b}. The apparitions needing the least depth of material were either Phaethon, the largest low-$q$ object in our sample, or objects in extreme forward-scattering geometries (${\alpha}>160^\circ$). 

Phaethon has been repeatedly detected in STEREO images at peak apparent magnitude of $\sim$11 \citep{hui17}. As this is about $\sim$3 mag fainter than SOHO's limiting magnitude, it is no wonder that it has never been detected in SOHO images. Thus, it seems likely that our simple model is overly optimistic. There are numerous uncertainties in our estimates, including assuming spherical grains, uniform grain size, grain albedo, and that asteroid grains scatter in a manner analogous to comet grains, which might affect the estimated depth by an order of magnitude. Furthermore, we assumed that such dust would be present throughout the time the asteroid was in SOHO's field of view, but \citet{li13b} showed that Phaethon's activity only lasts for a few days with peak activity $\sim$0.5 days after perihelion, greatly reducing the window for such dust to be detected on a given apparition. Thus, we can only exclude activity in the rest of the low-$q$ population at 1--3 orders of magnitude higher mass-loss rates per unit surface area than Phaethon since they are significantly smaller. A deeper search of the SOHO data via shifting and stacking at the ephemeris rate \citep[e.g.,][]{hui19} could set upper limits for activity near perihelion for each object in the sample, but is beyond the scope of the current paper.

\subsection{Rotation Periods}
\label{sec:discussion_rot}

\begin{figure}[t!]
    \centering
    \includegraphics[width=0.5\textwidth]{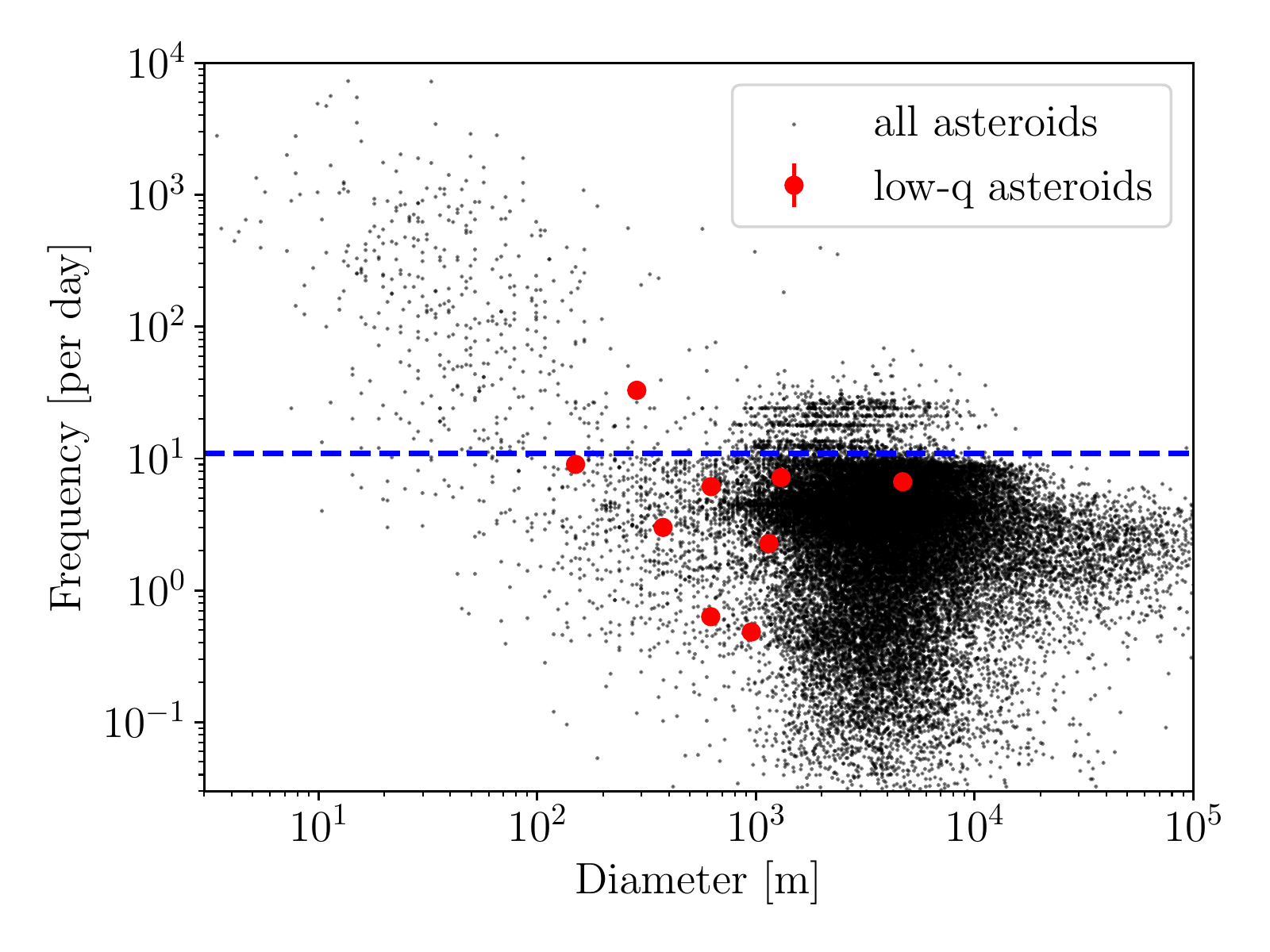}
    \caption{The distribution of rotational periods vs.\ diameter of \lowq asteroids (red) compared to the general asteroid population from the JPL Small-Body Database (black). The critical spin limit of 2.2 hours can be seen for asteroids with diameters greater than 200 meters. }
    \label{fig:rotation}
\end{figure}

Only nine \lowq objects have reported rotational periods, including 2006 HY51 first reported in this work (Table \ref{tab:published}; Figure \ref{fig:rotation}). While several \lowq objects have rotation periods $\sim$3--4 hours, only one object (2011 XA3) has a rotation period of $\sim$45 minutes, which is below the critical spin limit for a rubble pile asteroid \citep{urakawa14}. 2011 XA3 has $H=20.4$, corresponding to a diameter $\sim$200 -- 500 meters, depending on albedo. Because fast-rotating asteroids must resist their own centrifugal force, they have structurally significant tensile strength.

Both 322P and 96P are among the fastest-rotating known comets being at/near the spin period limit, with periods of $\sim$3 and $\sim$4 hours respectively \citep{knight16,eisner19}. 323P is the fastest rotating comet with a period of $\sim$0.5~hrs, which is a suggested driver of the comet's activity \citep{hui22}. The fast spin-rates could be indicative of increased YORP effect near the Sun. A larger sample in the future could support this idea if more \lowq objects were found to be fast-rotating.

Smaller asteroids are found to have bluer colors \citep{binzel04,thomas12,carry16}, likely due to resurfacing caused by YORP spin-up and failure \citep{graves18}. Therefore, we might assume faster-spinning objects approaching or above the spin barrier have bluer colors. On the other hand, a faster-spinning asteroid has a more smoothed-out temperature distribution in longitude than a slower rotating one, creating less diurnal variability and lessening the effects of resurfacing due to thermal fatigue. We search for a trend of colors versus rotational period. However, the sample is very limited (nine objects) and no definitive conclusion can be made.

\subsection{Future Surveys}
The upcoming Vera Rubin Observatory's Legacy Survey of Space and Time (LSST) will increase the number of known NEOs by an order of magnitude. NEOs will be observed a median of 90 times spread among \textit{ugrizy} bandpasses \citep{jones16}. A similar analysis to this work with a much larger sample size will likely be possible. However, if different filters are not obtained simultaneously, colors might not be easily measured due to geometric and rotational variation between observations.  Because there will be hours to days between observations of the same object, rotational periods are unlikely to be determined without dedicated prompt follow-up of LSST discoveries, unless the period is sufficiently long. LSST will observe NEOs at several epochs with various phase angles so a phase curve is likely to be obtained.  Determining the phase curve for \lowq objects will give us a better understanding of surface roughness, which can better constrain the effects of regolith production and resurfacing due to thermal fatigue. LSST will be capable of monitoring and reporting any activity or disruption events nightly with potential pipelines and proposed surveys \citep{schwamb18,seaman18}, which will further the study of near-Sun asteroids that are more likely to be active or catastrophically disrupted \citep{jewitt12,granvik16,ye19}. 

The Near-Earth Object Surveillance Mission \citep[NEOSM, previously NEOCam;][]{sonnett19} aims to extend the Wide-field Infrared Survey Explorer (WISE) catalog of diameters and visual albedos by an order of magnitude. NEOSM is expected to point closer to the Sun than NEOWISE or Spitzer, enabling the search for near-Sun activity as well as the discovery and albedo measurements of more \lowq objects. More albedo measurements of \lowq asteroids will help us determine the surface temperature these objects reach as well as how space weathering affects the surface, which varies with composition and can be distinguished with albedo \citep{lantz17}. Additionally, the combination of NEOSM with increasing cadence of visible observations like LSST (and other surveys like PANSTARRS, ZTF) should improve the size estimates.

\section{Conclusions}
\label{sec:conclusions}
We obtained magnitude and optical color measurements of 22 near-Sun asteroids with $q\leq0.15$ au over three years from January 2017 to March 2020 primarily using various 4-m class telescopes. We obtained lightcurve data for three \lowq asteroids, finding results consistent with previously published values for two and determining a rotation period for 2006 HY51 equal to $3.350\pm0.008$\,hr. We provide a summary of all known properties of near-Sun asteroids including albedos, rotational periods, and spectra. These objects were studied in order to search for trends relating to surface  modification due to near-Sun processes, particularly those that might lead to disruption. We find that the observed \lowq asteroids exhibit bluer colors overall, though overlapping with the color distributions of NEAs. However, there are no clear trends of colors with perihelion distance, Tisserand parameter, or rotational period. Unknown dynamical histories and compositions for individual objects combined with competing surface altering processes are likely responsible for the stochastic color distribution.  Finally, future surveys will enable studies of near-Sun objects with a much larger sample.

\section*{ACKNOWLEDGEMENTS}
We thank the anonymous referees for their thoughtful review. CH would like to thank Mikael Granvik and Robert Jedicke for assistance in generating source regions for our sample, Richard Binzel for thoughtful ideas, and Anathasia Toliou for assistance interpreting the dwell times results. CH, MMK, MSK, and QY are supported by NASA Near Earth Object Observations grant NNX17AH06G. We thank Lori Feaga, Tony Farnham, and Nick Moskovitz for assisting with some observations and Brian Skiff and Larry Wasserman for their assistance in scheduling observations using Lowell Observatory's 31-inch telescope.

These results made use of the Lowell Discovery Telescope at Lowell Observatory. Lowell is a private, non-profit institution dedicated to astrophysical research and public appreciation of astronomy and operates the LDT in partnership with Boston University, the University of Maryland, the University of Toledo, Northern Arizona University, and Yale University. The Large Monolithic Imager was built by Lowell Observatory using funds provided by the National Science Foundation (AST-1005313). 

These results are based in part on observations obtained at the Southern Astrophysical Research (SOAR) telescope, which is a joint project of the Minist\'{e}rio da Ci\^{e}ncia, Tecnologia e Inova\c{c}\~{o}es (MCTI/LNA) do Brasil, the US National Science Foundation's NOIRLab, the University of North Carolina at Chapel Hill (UNC), and Michigan State University (MSU).

The INT is operated on the island of La Palma by the Isaac Newton Group of Telescopes in the Spanish Observatorio del Roque de los Muchachos of the Instituto de Astrofísica de Canarias. INT data were obtained under program I/2016B/02.

The Pan-STARRS1 Surveys (PS1) and the PS1 public science archive have been made possible through contributions by the Institute for Astronomy, the University of Hawaii, the Pan-STARRS Project Office, the Max-Planck Society and its participating institutes, the Max Planck Institute for Astronomy, Heidelberg, the Max Planck Institute for Extraterrestrial Physics, Garching, the Johns Hopkins University, Durham University, the University of Edinburgh, the Queen's University Belfast, the Harvard-Smithsonian Center for Astrophysics, the Las Cumbres Observatory Global Telescope Network Incorporated, the National Central University of Taiwan, the Space Telescope Science Institute, the National Aeronautics and Space Administration under Grant No.\ NNX08AR22G issued through the Planetary Science Division of the NASA Science Mission Directorate, the National Science Foundation Grant No.\ AST-1238877, the University of Maryland, Eotvos Lorand University (ELTE), the Los Alamos National Laboratory, and the Gordon and Betty Moore Foundation.

Funding for the SDSS and SDSS-II has been provided by the Alfred P. Sloan Foundation, the Participating Institutions, the National Science Foundation, the U.S. Department of Energy, the National Aeronautics and Space Administration, the Japanese Monbukagakusho, the Max Planck Society, and the Higher Education Funding Council for England. The SDSS Web Site is http:// www.sdss.org/.

The SDSS is managed by the Astrophysical Research Consortium (ARC) for the Participating Institutions. The Participating Institutions are the University of Chicago, Fermilab, the Institute for Advanced Study, the Japan Participation Group, The Johns Hopkins University, the Korean Scientist Group, Los Alamos National Laboratory, the Max-Planck-Institute for Astronomy (MPIA), the Max-Planck-Institute for Astrophysics (MPA), New Mexico State University, University of Pittsburgh, University of Portsmouth, Princeton University, the United States Naval Observatory, and the University of Washington.

%

\vspace{5mm}
\facilities{DCT, SOAR, ING:Newton, LO:0.8m, Hall}


\software{{\tt astropy} \citep{astropy18}, {\tt astroquery} \citep{astroquery19}, {\tt calviavat} \citep{kelley19}, {\tt ccdproc} \citep{ccdproc}, JPL Horizons \citep{horizons}, NEOPOP\footnote{https://neo.ssa.esa.int/neo-population-generator} \citep{granvik18b}, {\tt PHOTOMETRYPIPELINE} \citep{mommert17}, {\tt photutils} \citep{photutils_1.2.0}, {\tt seaborn} \citep{waskom21}
          }

\bibliography{allreferences}{}
\bibliographystyle{aasjournal}



\listofchanges

\end{CJK*}
\end{document}